\pdfoutput=1
\documentclass[11pt,a4paper]{article}

\usepackage[utf8]{inputenc}
\usepackage[T1]{fontenc}
\usepackage{lmodern}
\usepackage[margin=2.5cm]{geometry}
\usepackage{amsmath,amssymb}
\usepackage{graphicx}
\usepackage[table]{xcolor}
\usepackage{array}
\usepackage{tabularx}
\usepackage{booktabs}
\usepackage{float}
\usepackage{enumitem}
\usepackage{pdflscape}   
\usepackage{siunitx}
\usepackage{authblk}
\usepackage[hidelinks]{hyperref}

\definecolor{AccentBlue}{HTML}{2E2EFF}
\definecolor{main}{HTML}{2E2EFF}
\definecolor{sub}{HTML}{E6E6FF}
\definecolor{headergray}{RGB}{240,240,240}
\definecolor{rowline}{RGB}{220,220,220}
\definecolor{headerblue}{RGB}{45,85,170}
\definecolor{bodyblue}{RGB}{232,240,252}

\title{When AI Fails, What's Next? A Data-Driven Taxonomy of Real-World AI Incident Responses}

\author[1,*]{Evgenija Popchanovska}
\author[1]{Ana Gjorgjevikj}
\author[2,*]{Maryan Rizinski}
\author[2]{Lubomir T. Chitkushev}
\author[3]{Irena Vodenska}
\author[1,2]{Dimitar Trajanov}

\affil[1]{Faculty of Computer Science and Engineering, Ss. Cyril and Methodius University, 1000 Skopje, North Macedonia}
\affil[2]{Department of Computer Science, Metropolitan College, Boston University, Boston, MA 02215, USA}
\affil[3]{Department of Administrative Sciences, Metropolitan College, Boston University, Boston, MA 02215, USA}
\affil[*]{Corresponding authors: \href{mailto:evgenija.popchanovska@students.finki.ukim.mk}{evgenija.popchanovska@students.finki.ukim.mk}; \href{mailto:rizinski@bu.edu}{rizinski@bu.edu}}

\date{}

\begin{document}

\maketitle

\begin{abstract}
Large language models (LLMs) are now part of many high-stakes workflows. When they fail, the harm can go beyond the model and cause legal, reputational and financial damage. This paper describes the publicly reported responses of organizations and other stakeholders to such incidents. We use 9,705 incident entries from three repositories of publicly reported AI incidents: AIID, AIAAIC and the OECD AI Incidents Monitor (AIM). Each entry is represented by one media article. We extract mitigation actions with structured prompting and classify \textbf{6,893 entries}. We also propose an LLM-based pipeline for this task, which we expect to become better and more reliable over time as language models improve. Building upon MIT's AI Risk Mitigation Taxonomy (RMT), we propose the AI Risk Response Taxonomy (RRT). It adds four categories: Corrective \& Restrictive Actions; Legal, Regulatory \& Enforcement Actions; Financial, Economic \& Market Controls; and Avoidance \& Denial. The RRT has 32 subcategories in total. Grounded in socio-technical safety and crisis communication theory, it separates constructive risk controls from institutional posturing and contestation. The added subcategories account for 1,556 of 4,218 labels in AIID (36.9\%), 1,670 of 4,500 in AIAAIC (37.1\%) and 6,403 of 15,276 in the OECD AIM (41.9\%). These shares show that the added response types are common in media-reported AI incidents. At the record level, 76.9--80.5\% of the classified entries contain at least one added subcategory. The RRT establishes an analytical framework and shared vocabulary to help AI researchers, developers and policymakers better understand AI risks through the analysis and evaluation of publicly documented incident responses.
\end{abstract}


\section{Introduction}\label{introduction}

The field of artificial intelligence (AI) has transformed profoundly in the last decade, driven largely by advances in deep learning (DL) and the development of large-scale general-purpose models \cite{lecun2015deep, wang2025history, annepaka2025large}. Among these, large language models (LLMs) based on the transformer architecture have risen to prominence, driving a paradigm shift from narrow task-specific models to general-purpose systems \cite{vaswani2017attention, devlin2019bert, radford2018improving}. LLMs exhibit remarkable performance in a wide range of tasks that involve natural language understanding, generation and reasoning \cite{achiam2023gpt, bubeck2023sparks, wang2019superglue}. Instead of designing specialized models for each task, the current trend is to pretrain one general-purpose large model, then adapt it for many tasks \cite{bommasani2021opportunities, brown2020language}. Due to their capabilities, LLMs are rapidly being integrated across diverse industries, including software engineering, finance, healthcare, and law, promising gains in efficiency and decision-making. LLM-based systems are deployed in a variety of environments, from automated medical diagnosis to algorithmic trading and legal document analysis, all of which lead to a new era of human-computer collaboration.

However, this rapid and widespread adoption carries significant risks alongside its benefits. The complexity and opaque nature of modern AI systems, particularly LLMs, introduce a new set of systemic risks that extend far beyond simple performance errors \cite{bommasani2021opportunities, uuk2024taxonomy, hendrycks2023overview}. The deployment of inadequately vetted or poorly governed models may expose organizations to significant legal and financial liabilities that can be attributed to flawed advice and decision-making, copyright infringement, and data privacy violations, among others \cite{neel2023privacy, wachter2024large, carnat2024addressing}. Beyond financial losses, ethical issues such as bias, discrimination, and the potential for malicious use present profound challenges that are capable of perpetuating and even amplifying existing disparities \cite{ferrara2024fairness, ferrer2020bias, mehrabi2021survey, jui2024fairness}. Moreover, high-profile failures can cause severe reputational harm, leading to an erosion of public trust in AI technologies and the organizations that deploy them \cite{holweg2022reputational, prahl2021rogue}. This growing awareness of potential harms has spurred increased regulatory scrutiny globally, with policymakers developing legislation, standards and frameworks aimed at ensuring the safe, transparent and accountable use of AI \cite{de2021artificial, camilleri2024artificial, nastoska2025evaluating}.

Many prominent AI failures are not isolated incidents but are symptomatic of deeper, systemic weaknesses in the AI development and deployment lifecycle. While individual models may exhibit technical shortcomings, these weaknesses often extend well beyond the model level to encompass the surrounding data pipelines, system integration and deployment, operational practices, human oversight and organizational decision-making across the entire end-to-end system. These systemic failures often arise from a confluence of interconnected issues that are difficult to anticipate in traditional software engineering. However, while a growing body of literature, including news outlets, analyzes individual AI incidents, a significant knowledge gap persists. Much of the existing research provides descriptive accounts of specific failures or focuses on rather narrow technical solutions. What is largely missing is a structured, overarching framework that synthesizes insights from disparate failures to provide a systemic understanding of how organizations and other stakeholders respond when they occur. The field lacks a comprehensive taxonomy---based in real-world incidents rather than merely theoretical insights---that systematically describes the responses and mitigation actions actually reported after AI incidents across the entire AI lifecycle, from data curation and model design to post-deployment monitoring and governance. Without such a framework, organizations across industries are left in a reactive posture, addressing failures as they arise rather than proactively building resilient and trustworthy AI systems.

This paper aims to address this gap. We perform analysis of selected, high-impact cases of systemic AI failures in real-world systems, with a particular emphasis on incidents involving LLMs. Our objective is not merely to catalogue these events but to distill from them recurring patterns in how AI incidents are responded to. To this end, we construct a novel dataset of AI-related incidents grounded in real-world cases, drawing from the AI Incident Database (AIID), the AI Incidents Monitor (AIM) of the Organisation for Economic Co-operation and Development (OECD), and the repository of the AI, Algorithmic and Automation Incidents and Controversies (AIAAIC) initiative. We use this dataset to analyze the responses and mitigation actions reported in response to these incidents. Comprising more than 9,000 media articles, the resulting dataset integrates a substantially broader collection of AI incidents documented across diverse media sources. The code and dataset for this paper are available on GitHub at: \href{https://github.com/popchanovska/ai-systemic-failures}{github.com/popchanovska/ai-systemic-failures}.

Based on this approach, we propose a novel taxonomy of real-world AI incident responses, the AI Risk Response Taxonomy (RRT), building upon MIT's AI Risk Mitigation Taxonomy (RMT). This taxonomy serves as an analytical framework to help researchers, developers, and policymakers understand how AI incidents are publicly responded to in practice and to identify response types that existing normative frameworks do not capture. The primary contribution of this work is the explicit mapping of each reported response action to a specific category of the RRT. In addition, we trace the publicly reported responses in concrete use cases. The RRT is descriptive: it records which responses are reported in publicly available sources after AI incidents, not whether those responses reduced harm, prevented recurrence, or improved recovery. A reported response may be an announced commitment or advice to users; the RRT does not verify that an action was carried out. A reported restriction, lawsuit, compensation payment or organizational commitment therefore does not establish that the response was carried out as reported or that it was effective. Evaluating response effectiveness is left to future work, for which the RRT provides a structured starting point. Ultimately, we argue that fostering a culture of awareness around systemic failures is indispensable to harnessing the potential of AI while building production-grade AI-based systems.

The remainder of this paper is organized as follows. Section~\ref{literature_review} provides a review of state-of-the-art strategies to mitigate AI risks across the AI model lifecycle. Section~\ref{data_and_methodology} presents our methodology for constructing a novel dataset of high-profile cases of incidents involving AI-related risks, describing the complete pipeline from data collection and annotation, to incident structuring and classification. Section~\ref{data_and_methodology} also proposes the RRT by integrating insights from the dataset, with the aim of enabling a more comprehensive categorization of reported incident responses. Section~\ref{labeling} explains how the categorization was performed and presents the distribution of category labels across mitigation tactics. Section~\ref{impact_across_industries} compares five high-impact incident archetypes by tracing their reported responses through the taxonomy. Finally, Section~\ref{conclusion} concludes by summarizing our contributions and outlining directions for future research.

\section{Literature Review}\label{literature_review}

The rapid integration of large language models (LLMs) into critical infrastructure has shifted the research focus from isolated algorithmic errors towards systemic vulnerabilities. Traditional AI risk literature often categorizes failures into narrow technical silos, mainly focusing on model-level issues such as data bias or adversarial attacks. However, recent research suggests that systemic failure is an emergent property of the interaction between complex technical architectures and human-based governance. This section synthesizes current research and highlights its importance across several areas, including the evolution of AI failure frameworks, technical design flaws, governance gaps, high-stakes domains such as financial applications, trends in the taxonomy and categorization of AI risks and state-of-the-art mitigation strategies.

\textit{The Evolution of AI Failure Frameworks}. Early research into AI failures primarily focused on the reliability of narrow machine learning models. Seminal studies in the mid-2010s, such as \cite{amodei2016concrete} laid the groundwork by identifying unintended side effects and reward hacking as primary risks. As systems evolved into deep learning architectures, researchers began documenting the ``brittleness'' of neural networks when faced with out-of-distribution (OOD) data. However, as LLMs gained widespread prominence, the narrative shifted from simple model inaccuracy to systemic propagation, where a single model failure (e.g., a hallucination) triggers a cascade of legal and operational consequences across an enterprise ecosystem. Despite expanding industry safety frameworks, the emergence of sophisticated risks and the increasing complexity of pre-deployment testing present significant challenges for effective risk management in AI-based systems \cite{Bengio2026AISafety}. Such failures are not only technical. They are shaped by socio-technical feedback loops: how organizations monitor their systems, report problems, and react to external pressure determines whether a technical fault becomes an incident and how long its harm lasts.

\textit{Technical Design Flaws and Algorithmic Vulnerabilities}. A significant body of literature investigates the technical root causes of failure. A recurring theme is the ``black box'' nature of transformer architectures, which complicates model debugging and validation. Research findings and advances in Explainable AI (XAI) emphasize that a lack of interpretability is not merely a technical hurdle but a systemic risk that prevents human operators from intervening during a failure \cite{ribeiro2016should}. Hallucinations are another limitation of LLMs, arising from unclear knowledge boundaries, noisy training data, and memorized inaccuracies. Despite improved safeguards and model-editing techniques, LLMs remain susceptible to jailbreak prompts, posing risks in high-stakes domains such as healthcare, law, and finance. These failures erode decision-making and public trust, which emphasizes the need for more robust and responsible deployment \cite{zhang2025siren, zheng2025rsafe, diederich2025rule, yuan2024rigorllm}. Additionally, recent studies highlight that prompt injection and data poisoning have emerged as dominant technical threats, moving beyond theoretical possibilities to documented exploits in production environments \cite{alber2025medical, yi2025benchmarking, xiang2024badchain}. LLM-based systems further introduce risks such as privacy leakage, value misalignment, and toxic outputs, which require risk mitigation across the entire AI system lifecycle \cite{wang2025survey, feretzakis2024privacy, ullah2024privacy}.

\textit{Organizational Perspectives and Governance Gaps}. Modern research increasingly adopts a perspective on human impact, arguing that AI failures are rarely purely technical. Several studies argue that ``organizational silence'' and the ``responsibility gap'' are significant contributors to systemic risk \cite{santoni2021four, ali2023walking, kiener2022can, matthias2004responsibility}. Other papers such as \cite{mitchell2019model, d2022underspecification, gebru2021datasheets} have highlighted how the lack of rigorous documentation leads to downstream failures when models are applied in contexts for which they were never intended. This theme suggests that governance gaps are as detrimental to AI systems' integrity as model imperfections and errors.

\textit{High-Stakes Domains: The Financial Sector Case Study}. The financial industry serves as one of the most demanding evaluation domains for AI risk research. Various studies have explored how AI-based systems in finance introduce unique systemic risks, such as algorithmic collusion and market flash crashes, and that responsible use of AI algorithms is essential \cite{fish2024algorithmic, shabsigh2023generative, mcclellan2025ai, min2022systemic, rizinski2022ethically}. Unlike other applications, financial AI requires a higher threshold for ``groundedness'', meaning that model outputs must be tightly anchored to verifiable data, explicit assumptions, and real-world financial constraints to avoid spurious or destabilizing decisions. The research in \cite{wu2023bloomberggpt} on BloombergGPT and similar finance-oriented LLMs highlights the tension between the generative capabilities of these models and the rigid regulatory requirements of the financial sector. The consensus among financial AI researchers is that current validation practices, such as backtesting, are insufficient for the non-deterministic nature of generative agents \cite{fan2025ai, larooij2025validation}. The legal sector provides a well-documented record of how generative AI failures are handled after the fact. In \textit{Mata v.~Avianca, Inc.}, a US federal court sanctioned two attorneys and their law firm for filing a brief that cited non-existent cases generated by ChatGPT, and imposed a monetary penalty~\cite{mata2023avianca}. Empirical studies show that such failures are not isolated: general-purpose LLMs hallucinated in 58\% to 88\% of responses to verifiable questions about US federal court cases~\cite{dahl2024large}, and AI legal research tools from two major legal publishers each hallucinated in more than 17\% of queries~\cite{magesh2025hallucination}. This ruling and these studies anchor the legal cases analysed in Section~\ref{impact_across_industries}.

\textit{Taxonomy and Categorization Trends}. There is a growing debate in the literature regarding how to best categorize AI risks \cite{nastoska2025evaluating}. Traditional taxonomies, like the NIST AI Risk Management Framework (RMF), provide broad categories such as ``Valid \& Reliable'' or ``Fair -- with Harmful Bias Managed'' \cite{nist_ai_rmf}. However, recent papers argue that these frameworks are often too abstract for system architects. There is a visible trend toward more practical, multidimensional taxonomies that map specific technical failure modes (e.g., biased training data) to their negative impacts (e.g., systemic discrimination) \cite{slattery2025airiskrepositorycomprehensive, xia2023towards}. Our paper builds in part on this trend by proposing a taxonomy of reported incident responses that connects high-level policy categories with the concrete response actions reported in public sources, integrating insights from prominent existing taxonomies.

\textit{State-of-the-Art Mitigation Strategies}. Current state-of-the-art (SOTA) mitigation strategies have evolved from simple ``guardrails'' to complex multi-layered architectures. The strategies involve various approaches such as red teaming, Reinforcement Learning from Human Feedback (RLHF) and Retrieval-Augmented Generation (RAG), among others. Red teaming, which involves structured, adversarial testing of AI systems by simulating malicious, unexpected, or edge-case interactions across the full system stack, has become a standard approach for identifying failure modes in large language models (LLMs) prior to deployment \cite{ge2024mart, lee2024learning, deng2023attack, nother2025text}. While red teaming is effective at uncovering rare, high-impact vulnerabilities that are difficult to detect through conventional testing, it is resource-intensive and may fail to systematically cover broader classes of risks or emergent behaviors over time. Another technique is Reinforcement Learning from Human Feedback (RLHF)---it involves training a model to align its outputs with human preferences by using human-provided evaluations or rankings as a reward signal, which guides the model to produce responses that are more helpful, safe, or aligned with desired behavior \cite{kaufmann2024survey, chaudhari2025rlhf}. RLHF combines reinforcement learning with supervised fine-tuning to iteratively optimize the model based on these feedback signals. However, while widely adopted, studies such as \cite{casper2023open, dahlgren2025helpful} argue that RLHF may merely hide problematic behaviors rather than eliminate them, leading to a ``false sense of safety''. Finally, RAG is widely cited as a primary technical mitigation for hallucinations \cite{tonmoy2024comprehensive, zhang2025hallucination}. RAG is based on grounding (i.e. augmenting) LLM outputs in retrieved external information---typically by referencing an authoritative database outside the AI system---which reduces hallucinations compared with standalone generative models. RAG improves factual accuracy and traceability by anchoring model outputs to external sources, making it particularly useful in high-stakes or knowledge-intensive domains. However, its effectiveness depends heavily on the quality, coverage and timeliness of the retrieval corpus and thus it may introduce additional system complexity and latency.

\textit{Theoretical Foundations: Socio-Technical Safety and Crisis Communication}. Research on AI systems has only begun to draw on the system safety literature that defines modern incident analysis~\cite{dobbe2022system,rismani2023plane}, and crisis communication research has begun to examine how companies respond when AI fails~\cite{prahl2021rogue}. We use four frameworks from these two literatures to position the RRT.

\textit{Normal Accident Theory.} Perrow argues that in systems with high interactive complexity and tight coupling, small failures can interact in unexpected ways~\cite{perrow1984normal}. Accidents then become a normal property of the system rather than an exception. AI systems embedded in organisational workflows show both features: a model output can pass quickly through automated steps into decisions that are hard to reverse.

\textit{STAMP and STPA.} Leveson's System-Theoretic Accident Model and Processes (STAMP) treats safety as a control problem~\cite{leveson2004new,leveson2011engineering}. A system is modelled as a hierarchical control structure. Legislators and regulators constrain organisations, management constrains operations, and operators constrain the technical process. Each level issues control actions downwards and receives feedback upwards. Accidents occur when safety constraints are missing or not enforced, or when feedback is missing, wrong or delayed. System-Theoretic Process Analysis (STPA) applies this model to identify unsafe control actions before they occur~\cite{leveson2018stpa}. The RMT itself lists STPA as an example of risk management~\cite{saeri2025mappingairiskmitigations}.

\textit{FRAM.} Hollnagel's Functional Resonance Analysis Method (FRAM) offers a complementary view~\cite{hollnagel2012fram}. It models a system as coupled functions whose everyday performance varies. Failures can emerge when this normal variability combines, or resonates, across functions, even if no single component fails. For AI systems, this describes how a plausible but wrong output can pass through review, publication and decision steps that each work as intended.

\textit{Situational Crisis Communication Theory.} Coombs' Situational Crisis Communication Theory (SCCT) explains how organisations respond publicly to crises~\cite{coombs2007protecting}. The choice of response depends on how much responsibility stakeholders attribute to the organisation. SCCT groups responses into deny strategies (denial, attacking the accuser, scapegoating), diminish strategies (excuses and justifications) and rebuild strategies (apology and compensation).

\textit{Corporate defensiveness as systemic friction.} Together, these frameworks clarify the role of corporate defensiveness after an AI incident. In SCCT terms, denial is a deny strategy: it protects reputation but does not address the hazard. In STAMP terms, it acts as friction in the control structure. It blocks or delays the feedback that regulators, courts and the public need, and it postpones the control actions that would reduce harm. Defensive responses are therefore not a weak or extended form of mitigation. They are responses that can prolong exposure to harm. This distinction underlies the dual-track model introduced in Section~\ref{dualtrack}.

\textit{Comparative Analysis of Major AI Risk Taxonomies and Identifying the Research Gap}. Existing AI risk taxonomies have made contributions in categorizing the multifaceted landscape of AI failures, yet they remain siloed by their specific objectives. The MIT AI Risk Repository \cite{slattery2025airiskrepositorycomprehensive} excels in its comprehensive breadth, serving as a ``meta-framework'' that synthesizes over 1,600 risk formulations into a continuously updated database; however, its primary limitation lies in its theoretical focus on model developers, often overlooking the granular mitigation needs of smaller deployers and third-party users. In contrast, the AIAAIC taxonomy \cite{abercrombie2024collaborativehumancentredtaxonomyai} provides a human-centered perspective, utilizing over 2,000 real-world incidents to map harms that are often missed by technical frameworks---such as the ``accountability gap'' for stakeholders who never directly interacted with the system. While the AIAAIC is important in capturing empirical harm, it lacks the prescriptive technical depth required for architectural risk mitigation. Meanwhile, the OECD AI Monitor \cite{OECDairisks} bridges policy and practice by characterizing AI systems based on intergovernmental principles, offering high-level interoperability for global regulators. Yet, like its counterparts, the OECD framework does not organize the responses reported after incidents into a structured taxonomy of response types. This fragmentation highlights a research gap that this paper aims to address: the need for an incident-informed taxonomy that describes how organizations and other stakeholders are publicly reported to respond to AI incidents, complementing existing theoretical mitigation frameworks. Such a taxonomy does not by itself establish which interventions are effective, but it provides the structured description of real-world responses on which later effectiveness evaluations can build.

\section{Data and Methodology}\label{data_and_methodology}

The following sections describe the data and methodology we use to build upon MIT's AI Risk Mitigation Taxonomy (RMT) and to reflect real-world incident responses. Table \ref{tab1} presents the original taxonomy alongside our extension, which introduces four new categories and nine subcategories; the additions are highlighted in blue. Because it incorporates newly observed response actions from real-world incidents, the resulting taxonomy is hereafter referred to as the AI Risk Response Taxonomy (RRT).

The RMT is created by collecting 831 mitigation measures from 13 existing frameworks into a database, then grouping them into four main categories with 23 subcategories total. The taxonomy is organized into the following four categories:

\begin{itemize}
    \item[] 1. Governance \& Oversight Controls
    \item[] 2. Technical \& Security Controls
    \item[] 3. Operational Process Controls
    \item[] 4. Transparency \& Accountability Controls
\end{itemize}

The taxonomy is developed with three primary goals: to organize similar mitigations into coherent groups, to ensure the categories are clear and accessible to diverse audiences, and to encompass a broad range of possible mitigations \cite{saeri2025mappingairiskmitigations}. It is limited to approaches that organize mitigations based on the AI system lifecycle and the target of the intervention. Specifically, mitigations are clustered according to stages of AI development, deployment and use (such as design, training, testing and deployment), and by whether the intervention addresses human or organizational behavior. 

\textbf{Motivation for the research.} However, the RMT adopts a theoretical approach, deriving taxonomic features solely from published literature rather than from analysis of real-world AI incidents. 
A taxonomy grounded in empirical incident data is better positioned to capture the full range of responses that organisations are publicly reported to take, including response types that do not appear in normative frameworks. For this reason, we analyze the responses reported in real AI incidents and build upon the RMT with new categories and subcategories derived from real-world use cases in order to strengthen the framework. This retrospective approach increases the current framework's value by first identifying risks and the associated mitigation measures taken in response---and then systematically classifying them. The methodological choice is motivated by, and aligned with, MIT's call for a systematic mapping between AI risks and mitigation measures \cite{saeri2025mappingairiskmitigations}. In addition to introducing new mitigation actions, we identified and classified denial and defensive response actions observed in real-world AI incidents. As a result, the RRT builds upon the RMT by also covering responses that are not mitigations in the conventional sense. The purpose of this extension was to capture the reactions and practical measures undertaken by companies and other stakeholders following AI failures. By incorporating these real-world responses, the extended taxonomy provides broader coverage of the actions commonly implemented when AI-related incidents occur.

\begin{table}[H]
\caption{AI Risk Response Taxonomy (RRT), which builds upon the RMT by adding four new categories and nine subcategories. Newly added entries are highlighted in blue.}\label{tab1}%
\centering
\renewcommand{\arraystretch}{1.25}
\setlength{\tabcolsep}{6pt}

\definecolor{headergray}{RGB}{57, 62, 70}
\definecolor{bodygray}{RGB}{252,252,253}
\definecolor{headerblue}{RGB}{35,75,150}
\definecolor{bodyblue}{RGB}{215,222,242}

\begin{tabular}{>{\columncolor{bodygray}}p{0.47\linewidth}
                >{\columncolor{bodygray}}p{0.47\linewidth}}
\hline
\rowcolor{headergray}
\color{white}\textbf{1. Governance and Oversight Controls} & 
\color{white}\textbf{2. Technical and Security Controls} \\
\hline
1.1 Board Structure and Oversight & 2.1 Model and Infrastructure Security \\
\hline
1.2 Risk Management & 2.2 Model Alignment \\
\hline
1.3 Conflict of Interest Protections & 2.3 Model Safety Engineering \\
\hline
1.4 Whistleblower Reporting and Protection & 2.4 Content Safety Controls \\
\hline
1.5 Safety Decision Frameworks & \\
\hline
1.6 Environmental Impact Management & \\
\hline
1.7 Societal Impact Assessment &  \\
\hline
\rowcolor{headergray}
\color{white}\textbf{3. Operational Process Controls} & 
\color{white}\textbf{4. Transparency and Accountability Controls} \\
\hline
3.1 Testing and Auditing & 4.1 System Documentation \\
\hline
3.2 Data Governance & 4.2 Risk Disclosure \\
\hline
3.3 Access Management & 4.3 Incident Reporting \\
\hline
3.4 Staged Deployment & 4.4 Governance Disclosure \\
\hline
3.5 Post-Deployment Monitoring & 4.5 Third-Party System Access \\
\hline
3.6 Incident Response and Recovery & 4.6 User Rights and Recourse \\
\hline
\rowcolor{bodyblue}
3.7 Incident Investigation & 4.7 Training and Supportive Measures \\
\hline
\rowcolor{headerblue}
\color{white}\textbf{5. Corrective and Restrictive Actions} & \color{white}\textbf{6. Legal, Regulatory, and Enforcement Actions} \\
\hline
\rowcolor{bodyblue}
5.1 System and Feature Restrictions & 6.1 Court and Law Enforcement Interventions \\
\hline
\rowcolor{bodyblue}
5.2 Usage and Access Limitations & 6.2 Regulatory, Policy and Legal Mandates \\
\hline
\rowcolor{headerblue}
\color{white}\textbf{7. Financial, Economic, and Market Controls} & \color{white}\textbf{8. Avoidance and Denial} \\
\hline
\rowcolor{bodyblue}
7.1 Financial, Economic, and Compensation Remedies & 8.1 Denial and Defensive-Based Actions \\
\hline
\rowcolor{bodyblue}
7.2 Market Access and Commercial Restrictions &  \\
\hline
\end{tabular}
\end{table}

\subsection{Data Collection}

This section describes the three datasets used to build the RRT upon the RMT. All datasets consist of AI-related incidents reported in publicly available news media. Each dataset follows a similar collection methodology, where every reported AI incident is recorded as a distinct entry and linked to its original source.

The datasets differ in their definitions of key concepts such as AI harm, risk, incident, and error. These conceptual differences reflect variations across the literature and may introduce inconsistencies when datasets are compared directly. To address this, we manually reviewed the documentation of each dataset, including schema definitions, contextual descriptions, and the taxonomies proposed by the respective projects across their dimensions. Despite these definitional differences, all three datasets document actual or potential harm caused by AI systems. For this reason, we analyze them collectively and treat them as a unified corpus that captures a broad spectrum of AI-related harms and incident types. The datasets used in this study are described below.

\begin{itemize}
    \item \textbf{AI Incidents Monitor (AIM).} The AI Incidents Monitor \cite{OECDairisks}, developed by the OECD, is a platform for collecting and analyzing AI incidents and hazards reported in global news media. The dataset is updated in near real time and is designed to support policymakers, AI practitioners, and other stakeholders in understanding emerging AI risks. Each incident is documented using a structured framework that includes metadata such as title, description, and sources, as well as detailed information on harms, impacts on people and the environment, economic context, data and input characteristics, AI model properties, task and output descriptions, and recorded mitigation actions.

    \item \textbf{AI Incident Database (AIID).} The AI Incident Database \cite{AIID_website} is constructed through systematic collection and aggregation of reports describing real-world AI failures and harms. Its goal is to make AI risks visible by documenting what can go wrong when intelligent systems are deployed across domains such as autonomous robots, language and vision models, autonomous driving, recommender systems, identification technologies, AI oversight systems, and healthcare applications. The database employs standardized identifiers inspired by the Common Vulnerabilities and Exposures (CVE) database and the National Transportation Safety Board Aviation Accident database. AIID is also used by the MIT AI Incident Tracker project \cite{MIT_AI_Incident_Tracker} to classify incidents according to risk and harm severity.

    \item \textbf{AIAAIC Repository.} The AIAAIC Repository \cite{aiaaic_repository} collects and documents incidents involving AI, algorithmic, and automation systems. It focuses on identifying, analyzing, and exposing cases where such systems cause harm or are misused, with particular attention to societal impact, transparency, and accountability. Incidents are structured using attributes such as time of occurrence, deployer and developer, system name, technology and intended purpose, taxonomies of news triggers, ethical issues, external harms, consequences, and responses. Each entry also includes a narrative summary and links to the original sources.
\end{itemize}

For each incident, the datasets retain the original text from the news source along with a link to the corresponding article. Across all three datasets, this results in a total of \textbf{9,705} AI incident news articles. \textbf{Unit of analysis.} Each of the 9,705 observations is one incident entry in one repository (AIID: 1,200; AIAAIC: 1,478; OECD AIM: 7,027), represented by a single news article: the first report linked to the incident in AIID, the first linked source in AIAAIC, and the first linked article in the OECD AIM. Further articles about the same incident within a repository were not included, so heavily reported incidents do not contribute additional records within a repository. The same incident may, however, be recorded in more than one repository, and a repository may occasionally list the same event twice. Cross-repository duplicates were not removed before extraction and classification, so an incident recorded in several repositories can contribute more than one record. A very small fraction of entries may contain inconsistencies in the extracted article text due to unsupported websites, inaccessible sources, or removed articles. Such entries yield no extracted actions and are excluded at the extraction step (Section~\ref{labeling}, Table~\ref{tab:flow}); because this exclusion is a model judgement, a small number of entries with reported responses may also be excluded.

\textbf{Scope: publicly documented responses.} Because every entry is based on a publicly available news article, the corpus documents what organizations publicly communicate about an AI incident, or what courts, regulators and journalists make visible, and not everything that organizations do. Much of the response to an AI incident takes place internally: engineering fixes, retraining and fine-tuning, red-teaming, changes to monitoring, and internal governance decisions are usually recorded in proprietary telemetry and internal documents that are not published. Legal action, regulatory orders, financial settlements and explicit public statements are, by contrast, more often public and more likely to be reported. The RRT is therefore an empirical taxonomy of publicly documented, media-visible AI incident responses. It can serve as an audit framework for the visible socio-technical friction around AI incidents, but it is not a census of the closed-door patch and remediation cycles inside organizations. The implications of this visibility bias for the label distribution are discussed in Section~\ref{labeling}.

To check that the repository data matched our research objectives, and to see how incidents are represented in individual articles, we also built a validation cohort. We combined automated web scraping, using a Python script with terms such as ``LLMs'', ``hallucinations'', ``large language models'' and ``failures'', with targeted manual searches that paired platform names (ChatGPT, Gemini, Perplexity) with terms such as ``mistake'', ``failure'' and ``incident''. Sources included The Verge, The Guardian, The New York Times, Reuters, Wired, Forbes and the NY Post. Of about 200 candidate articles, we kept the 50 that describe clear, verifiable failures of LLMs or generative AI tools. Because of these keywords, the cohort is skewed towards LLM and generative AI incidents. This reflects the retrieval design, not a restriction of the study's scope, and the cohort is suited to validating patterns in LLM-related incidents rather than representing all AI failures.

\textit{Validation cohort as a held-out set.} The 50 validation articles were not used to construct the taxonomy, so we analysed them as a held-out set. The RRT categories were sufficient to describe the responses reported in these articles, and the added categories materially enriched the description obtained with the RMT subcategories alone, supporting the taxonomy's applicability outside the three repositories. Because the cohort was selected by keyword search and manual review rather than random sampling, and several of its incidents are also recorded in the repositories, this is supportive rather than conclusive evidence of generalisation.

\subsection{Methodology}

This section explains the methodology used to develop the RRT using the AI incident datasets described in the preceding section. The methodology comprises three sequential steps: the extraction of mitigation actions from individual incident texts, the derivation of draft taxonomies from the extracted data in processing batches, and the manual consolidation and finalization of the taxonomy. Each step was applied independently to each constituent dataset, preserving the analytical boundaries between sources while ensuring methodological consistency across the entire corpus. The rationale for maintaining separate processing pipelines for each dataset was to allow examination of each source's unique distributional characteristics before integrating findings at the taxonomy level.

    \subsubsection{Step I: Extraction of mitigation actions from each incident} The first step involved the automated extraction of mitigation actions from 9,705 AI incident entries using prompt-engineered queries submitted to the GPT-5-mini language model via the OpenAI Batch API. Of the total corpus, mitigation actions were successfully identified in 6,902 texts; 6,893 of these were saved in the final classification table (Table~\ref{tab:flow}, Section~\ref{labeling}). The remaining texts were determined by the model to contain no explicit reference to mitigation or corrective measures, and were assigned a default label of ``No mitigation taken''. It is important to note that this absence label itself represents a model judgment rather than a confirmed empirical fact: a text may describe mitigation indirectly or in terms that fall outside the model's extraction logic, without the model recognizing it as such. 

    \textbf{Prompt design.} The design of the extraction prompt was not a single fixed decision but an iterative methodological process that evolved through three successive versions. Each version was qualitatively evaluated by the authors on a random sample of 50 mitigation actions through inspection of representative extraction outputs and refined in response to observed failure modes.

    The \textbf{first prompt} version framed the task broadly, instructing the model to identify any corrective or mitigation actions taken in response to an AI incident, and explicitly counted a wide range of response types including statements of intent, plans, apologies, warnings, and condemnations as valid mitigations. While this formulation successfully captured a wide variety of response actions, inspection of outputs revealed several systematic limitations. Most notably, the broad scope definition led the model to extract content that was not strictly post-incident in nature: preventive measures, background contextual information, and general organizational policies with no clear causal link to the described incident were frequently included alongside genuine reactive responses. Additionally, the prompt allowed the model to aggregate multiple distinct actions into a single composite sentence, reducing granularity and hindering downstream classification. The output formatting was also inconsistent across incidents, with some responses departing from the specified list-of-strings structure.

    The \textbf{second prompt} version introduced several targeted improvements. The extraction scope was narrowed to actions taken in direct response to harm, risk, or failure caused by an AI model or system, thereby excluding background policies and preventive measures not triggered by the incident in question. The instruction to treat each incident independently was retained. Formatting requirements were strengthened, and the model was more explicitly instructed to extract only actions stated in the text rather than inferred. Evaluation of outputs under this version showed improved precision, with fewer non-incident-linked actions appearing in the extracted lists. However, some ambiguity remained regarding the temporal boundary between preventive and reactive actions, and the model continued to occasionally conflate multiple discrete actions.

    The \textbf{third and final prompt} version resolved these remaining issues through a combination of definitional precision and stricter output constraints. The system prompt introduced an explicit temporal boundary, defining a mitigation action as any step taken after the incident to respond to, contain, investigate, remediate, reduce harm, acknowledge responsibility, or address consequences caused by an AI system. Preventive measures preceding the incident were explicitly excluded. The instruction to split co-occurring actions into separate items was reinforced. Output format requirements were tightened to mandate a valid JSON array of strings with no explanatory preamble, commentary, or markdown formatting. The final prompt also restricted extraction to AI-related incidents, instructing the model to return ``No mitigation taken'' if the source text did not describe an actual AI incident. This restriction reduced noise from texts that described non-AI events which had been included in the corpus.
    
    All system and user prompts are provided in Appendix ~\ref{secB1}.

    \textbf{Validation of the extraction.} To check the extraction, we drew a simple random sample of 100 of the 9,705 incident texts, matching the corpus composition, which contained 375 extracted action statements; an independent LLM judge (Claude Opus 5.5) compared each statement and each ``No mitigation taken'' output with the full source article. Overall, 86.7\% of the statements (325/375; 95\% Wilson CI 83--90\%) were faithful to the source, and all 26 ``No mitigation taken'' outputs were correct. Most unfaithful statements dropped the source's hedging or attribution rather than misstating the action, and errors were concentrated in non-English articles.

    \textbf{Methodological limitation.} The mitigation extraction process produces LLM-generated standardized representations of mitigation actions rather than direct textual quotations from source articles. As both the presence and absence of mitigation records are determined through model interpretation and prompt design, ambiguously described or implicit actions may be omitted or inconsistently classified, potentially propagating errors into later stages of taxonomy construction.

    \textbf{Example generated from the final prompt.} The following text presents an example of an AI incident from the AIID database, identified by its object ID, alongside the mitigation actions extracted using the prompt described above.

    \noindent
    \begin{minipage}{\linewidth}
    \vspace{1em}
    \hrule height 0.4pt
    \vspace{1em}

    \begin{tabbing}
    \hspace{4cm} \= \kill
    Object ID \> ObjectId(625763db343edc875fe639ff) \\
    Title \> YouTube Kids app is STILL showing disturbing videos \\
    Source domain \> dailymail.co.uk \\
    Extracted mitigation \>
    \parbox[t]{0.65\linewidth}{%
    'YouTube apologised for the disturbing videos on the YouTube Kids app.', 
    'YouTube removed more than 50 user channels.', 
    'YouTube stopped running ads on more than 3.5 million videos since June.', 
    'YouTube tightened enforcement of its Community Guidelines.', 'YouTube age-gated content that inappropriately targets families.', 'YouTube removed inappropriate content from the YouTube Kids app.', 'YouTube stated that when it discovers inappropriate content it quickly takes action to remove it from its platform.', 'YouTube said thousands of people will be working around the clock to flag content.', 'YouTube began using machine learning to identify the most harmful content, which is then automatically reviewed.'
    }
    \end{tabbing}
    \vspace{0.5em}
    \hrule height 0.4pt
    \end{minipage}
    \vspace{1em}

    \subsubsection{Step II: Deriving draft taxonomy from data batches} After extracting mitigation actions from the articles' textual content, we use the GPT-5-mini model to construct a taxonomy. For this stage, only a single prompt was used, since the goal was not to automatically create a final taxonomy, but to generate initial categories and terms to support later author-driven taxonomy development. We define both a system prompt and a user prompt for this task. A snippet of the system prompt used is: \\ 

    \noindent
    \begin{minipage}{\linewidth}
    \par\setlength{\leftskip}{2em}
    ``You are an AI incidents analyst specializing in qualitative analysis, clustering and AI Risk Response Taxonomy. Your task is to analyze a list of mitigation statements and derive a hierarchical taxonomy consisting of categories and subcategories based solely on the data. The taxonomy must be data-driven and generalizable.'' \\
    \par
    \end{minipage}

    The taxonomy derivation is performed in batches of the mitigation actions of 2,000 incidents each, so that the model can process the input within token limits. For each batch, there is a separate derived taxonomy. The prompt specifies that the output must follow a structured JSON format. 
    The following JSON snippet presents one category of the derived taxonomy from the AIID database. 

    \noindent
    \begin{minipage}{\linewidth}
    \vspace{1em}
    \hrule height 0.4pt
    \vspace{1em}

    \ttfamily\small
    \obeylines
    "derived taxonomy": \{
        \hspace*{2em}"Service and feature shutdowns or rollbacks": [
        \hspace*{4em}"Take demo/model/service offline or disable access",
        \hspace*{4em}"Limit or phase rollout of problematic features",
        \hspace*{4em}"Remove apps or features from stores/platforms",
        \hspace*{4em}"Shut down specific bots/services"
        \hspace*{2em}],
    \vspace{1em}
    \hrule height 0.4pt
    \end{minipage}
    \vspace{1em}

    \subsubsection{Step III: Constructing and finalizing the taxonomy} After obtaining the derived taxonomies from the GPT-5-mini model, we manually reviewed them by reading each category and subcategory and checking whether they match the RMT. Subcategories that did not match were stored separately. Once all unmatched subcategories were collected, we grouped them into clusters and defined new categories where appropriate. Through this process, we identified two new subcategories within existing categories: Incident Investigation and Training \& Supportive Measures. The Incident Investigation subcategory was added to the Operational Process Controls category, while the Training \& Supportive Measures subcategory was added to the Transparency \& Accountability Controls category. Furthermore, based on the nature of the remaining subcategories, we identified four new categories that are not represented as separate categories in the RMT, although several of their subcategories partly overlap with existing RMT subcategories (see Table~\ref{tab:crosswalk}): Corrective \& Restrictive Actions, Legal, Regulatory \& Enforcement Actions and Financial, Economic \& Market Controls and responsive Avoidance \& Denial based actions. The newly added categories and subcategories are shown in Table~\ref{tab1} and are marked blue.

    \textbf{Category boundary decision rules.} Because certain incident responses could plausibly be assigned to more than one category, we established the following explicit decision rules to define the intended category boundaries at the boundaries most likely to produce overlap. They guided the construction of the taxonomy but were not part of the classification prompt (Section~\ref{labeling}).

    \begin{enumerate}[label=\alph*.]
        \item \textit{Governance \& Oversight Controls (Category~1) vs. Legal, Regulatory \& Enforcement Actions (Category~6).} Category~1 captures \textit{internal} governance mechanisms voluntarily adopted by an organization (e.g., establishing an AI ethics board, updating internal risk management policies). Category~6 captures actions imposed by \textit{external} legal or regulatory authorities. When a company updates its governance structure in response to a regulatory mandate, the regulatory mandate is coded under Category~6, while any internal organizational changes that follow are coded under Category~1. When both dimensions are present in a single incident, multiple labels are assigned.

        \item \textit{Corrective \& Restrictive Actions (Category~5) vs. Operational Process Controls (Category~3).} Category~5 covers immediate, post-incident restrictions and rollbacks (e.g., disabling a feature, suspending a product deployment). Category~3 covers ongoing operational processes such as monitoring, auditing, and staged deployment. An action is coded under Category~5 when it is reactive and incident-triggered; it is coded under Category~3 when it reflects a planned or recurring operational practice.

        \item \textit{Financial, Economic \& Market Controls (Category~7) vs. Legal, Regulatory \& Enforcement Actions (Category~6).} Regulatory fines imposed by a government authority are coded under Category~6 (enforcement actions), while voluntary financial compensation to affected parties or civil lawsuit settlements are coded under Category~7.1. Court-ordered monetary damages are coded under both Subcategory 6.1 and Subcategory 7.1, as they arise from judicial proceedings while also constituting monetary compensation as part of the mitigation outcome.

        \item \textit{Avoidance \& Denial (Category~8).} This category is applied only when an actor \textit{explicitly and publicly} rejects responsibility, denies the existence of harm, or refuses remedial action, or when the source explicitly reports avoidance (non-response, material delay or non-disclosure of a known harm). It is not applied when an organization disputes specific technical claims while still implementing corrective measures. Where an actor simultaneously issues a public denial \textit{and} undertakes partial corrective action in response to the same incident, both Category~8 and the relevant remediation category are assigned to the respective actions independently.
    \end{enumerate}

    To further illustrate these rules, Table~\ref{tab:edge_cases} provides worked examples of edge cases and their assigned classifications.

    \begin{table}[H]
    \caption{Illustrative edge cases, including cases of partial overlap with RMT subcategories, and their taxonomy assignments under the category boundary decision rules.}
    \label{tab:edge_cases}
    \centering
    \renewcommand{\arraystretch}{1.3}
    \begin{tabularx}{\linewidth}{>{\raggedright\arraybackslash}X >{\raggedright\arraybackslash}p{3.5cm} >{\raggedright\arraybackslash}p{5cm}}
    \hline
    \textbf{Incident response} & \textbf{Category assigned} & \textbf{Rationale} \\
    \hline
    A national data protection authority fines a company \$10M and mandates a compliance audit & 6.2 + 7.1 (Regulatory Policy \& Legal Mandates + Financial, Economic \& Compensation Remedies) & External enforcement by regulatory body; monetary fine \\
    \hline
    A company voluntarily compensates affected users following a bias incident, without legal compulsion & 7.1 (Financial \& Compensation Remedies) & Voluntary monetary redress; no court or regulator involved \\
    \hline
    Following a regulatory investigation, a company also revises its internal AI governance charter & 6.2 + 1.2 (Risk Management) & External trigger coded under Category~6; resulting internal governance change coded under Category~1 \\
    \hline
    A company publicly states it complied with all applicable laws and declines to remove a system & 8.1 (Denial \& Defensive-Based Actions) & Explicit public refusal to remediate; legal defense invoked in lieu of corrective action \\
    \hline
    A company disables a chatbot feature after user complaints & 5.1 (System \& Feature Restrictions) & Reactive, incident-triggered product-level restriction \\
    \hline
    A company activates its pre-designed emergency rollback after a harmful output and coordinates the response & 5.1 + 3.6 (Incident Response \& Recovery) & Overlap with RMT 3.6, which already lists emergency shutdown and rollback procedures; the realised restriction is coded 5.1, the coordination of the response 3.6 \\
    \hline
    After a regulator's order, a company stops offering its model for a high-risk use in one country and updates its acceptable-use policy & 5.2 + 6.2 + 3.3 (Access Management) & Overlap with RMT 3.3, which covers high-stakes application prohibitions; the reactive, country-specific pause is coded 5.2, the order 6.2, the lasting policy change 3.3 \\
    \hline
    Following a chatbot error, a company conducts a root cause analysis and publishes its findings & 3.7 + 4.3 (Incident Reporting) & Overlap with RMT 3.1 and 4.3; the retrospective analysis of causes is coded 3.7, sharing the findings 4.3 \\
    \hline
    After users are harmed, a company opens a support line for affected users and retrains its staff & 4.7 + 4.6 (User Rights \& Recourse) & Overlap with RMT 4.6, which lists remediation protocols; support and training are coded 4.7, reporting and appeal channels 4.6 \\
    \hline
    \end{tabularx}
    \end{table}

    \textbf{Overlap with the RMT.} Some of the added subcategories make explicit, or divide into narrower types, concepts that the RMT already contains. For example, the RMT lists emergency shutdown and rollback procedures under Incident Response \& Recovery (3.6) and prohibitions on high-stakes applications under Access Management (3.3), which overlap with Subcategories~5.1 and~5.2. Like Category~5, Subcategory~3.6 applies after an incident, so the two are separated by what the action does rather than by its timing: Category~5 records restrictions of a system's capabilities or deployment scope, while 3.6 records containment, coordination and restoration (Section~\ref{taxonomy}). The RMT authors report similar overlap among their own subcategories~\cite{saeri2025mappingairiskmitigations}, and we do not claim that the additions are fully disjoint from the RMT. What they add is a related but distinct dimension. We therefore distinguish two kinds of additions. Categories~6, 7 and~8 add response dimensions that the RMT does not represent as separate controls: actions by external actors, namely courts, regulators and markets, and defensive responses by the responsible party. Category~5 and Subcategories~3.7 and~4.7 are finer subdivisions of existing RMT controls: Category~5 separates the restriction carried out after an incident from 3.6 and 3.3, Subcategory~3.7 separates the investigation of a specific incident from 3.6 and 3.1, and Subcategory~4.7 separates training and support measures from 1.1 and 4.6. The RMT describes prospective controls that an organization can plan and implement, organized by lifecycle stage and target of the intervention. The RRT records what was publicly reported as done after a specific incident and by whom. This includes actions by external actors such as courts, regulators, and markets, and responses such as denial that do not meet the RMT's definition of a mitigation as an action that reduces the likelihood or impact of a risk. Table~\ref{tab:crosswalk} maps each of the nine added subcategories to the closest RMT subcategories, states what the RRT adds, and gives an illustrative response.

    \begin{table}[!htbp] 
    \caption{Crosswalk between the nine subcategories added in the RRT and the closest subcategories of the RMT. RMT definitions and examples are abridged from~\cite{saeri2025mappingairiskmitigations}.}
    \label{tab:crosswalk}
        \centering
    \footnotesize
    \renewcommand{\arraystretch}{1.0}\setlength{\tabcolsep}{4pt}
    \begin{tabularx}{\linewidth}{>{\raggedright\arraybackslash}p{2.1cm} >{\raggedright\arraybackslash}X >{\raggedright\arraybackslash}X >{\raggedright\arraybackslash}X}
    \hline
    \textbf{RRT subcategory} & \textbf{Closest RMT subcategories (abridged definition)} & \textbf{What the RRT adds} & \textbf{Illustrative response} \\
    \hline
    3.7 Incident Investigation & 3.6 Incident Response \& Recovery: responds to incidents to contain harm and restore safe operations. 3.1 Testing \& Auditing: internal and external evaluations, including of compliance processes. & Retrospective analysis of the causes of a specific incident, as distinct from containment or routine evaluation & After its AI research team accidentally exposed internal data, Microsoft revoked the access token and completed an internal investigation into the potential impact (\texttt{AIID-00516}) \\
    \hline
    4.7 Training \& Supportive Measures & 1.1 Board Structure \& Oversight: lists crisis simulation training. 4.6 User Rights \& Recourse: reporting channels, appeals, remediation protocols. & Post-incident capacity building and support for workers, users, and affected people (e.g., retraining, more moderators, victim support) & Following a lawsuit over a teenager's suicide, Character.AI said it had hired a head of trust and safety, a head of content policy and additional engineering safety staff (\texttt{AIAAIC-00207}) \\
    \hline
    5.1 System \& Feature Restrictions & 3.6 Incident Response \& Recovery: lists emergency shutdown, rollback, and model containment. 2.3 Model Safety Engineering: lists capability restriction mechanisms. & The enacted, incident-triggered withdrawal or disabling of a deployed system or feature, recorded as a response rather than a planned procedure & After criticism of how it handles users in distress, OpenAI changed how ChatGPT answers questions on personal challenges and added reminders to take breaks (\texttt{OECD-05691}) \\
    \hline
    5.2 Usage \& Access Limitations & 3.3 Access Management: governs who uses AI systems and for what purposes, incl.\ high-stakes application prohibitions. 1.5 Safety Decision Frameworks: lists deployment pause triggers. & Reactive pause or restriction of an existing deployment in a specific region, sector, or context & OpenAI blocked access to ChatGPT in Italy after the national data protection authority's 2023 order (\texttt{AIID-00488})
    \\
    \hline
    6.1 Court \& Law Enforcement Interventions & 4.6 User Rights \& Recourse: enables affected users to seek recourse or remediation. & Litigation and actions by courts and law enforcement, i.e., actors outside the organization, which the RMT does not represent as separate controls & Two lawyers and their law firm were fined for submitting fabricated ChatGPT citations (\texttt{OECD-01332}); the parents of a teenager sued OpenAI (\texttt{AIID-01180})
    \\
    \hline
    6.2 Regulatory, Policy \& Legal Mandates & 1.2 Risk Management: lists compliance programs. 4.2 Risk Disclosure: pre-deployment notifications to government. 4.3 Incident Reporting: mandatory breach notification procedures. & The regulator's own action (investigation, order, fine), rather than the organization's compliance with it & Italy's data protection authority temporarily ordered OpenAI to stop processing Italian users' data (2023) (\texttt{AIID-00488})
    \\
    \hline
    7.1 Financial, Economic \& Compensation Remedies & 4.6 User Rights \& Recourse: lists remediation protocols, without specifying monetary redress. & Monetary responses: compensation, refunds, settlements, damages & Air Canada was ordered to compensate a passenger for a refund policy invented by its chatbot (\texttt{AIID-00552}) \\
    \hline
    7.2 Market Access \& Commercial Restrictions & 3.3 Access Management: API-only access, acceptable use policies. 3.4 Staged Deployment: limited API access programs. & Restrictions on market presence (listing, sale, procurement) rather than on operational use & Several technology companies announced moratoria on selling facial recognition systems to police in 2020 (\texttt{OECD-00040})
    \\
    \hline
    8.1 Denial \& Defensive-Based Actions & None. The RMT defines a mitigation as an action that reduces the likelihood or impact of a risk; denial is the opposite of 4.3 and 4.6. & Non-mitigative response behaviour, recorded for descriptive completeness (Section~\ref{taxonomy}) & A company executive's dismissive public reply to a protest (incident \texttt{OECD-04635}, Section~\ref{labeling}) \\
    \hline
    \end{tabularx}
        \end{table}

\subsection{Taxonomy}
\label{taxonomy}

This subsection presents a detailed examination of the new categories and subcategories added to the RMT to form the RRT. We have added four new categories (including their respective subcategories) and two additional subcategories to existing categories. The four newly introduced categories include the following:

\begin{enumerate}

\item \textbf{Corrective \& Restrictive Actions.} This category captures reactive, incident-triggered or risk-triggered actions taken to limit or remove AI system capabilities or deployment scope after a harm, failure, or identified risk has materialised. It is divided into two subcategories: System \& Feature Restrictions (5.1) and Usage \& Access Limitations (5.2). The distinction between the two lies in the \textit{target} of the restriction: 5.1 addresses constraints on the \textit{capabilities or features of the system itself}, whereas 5.2 addresses constraints on the \textit{deployment or use} of the system in specific contexts, regions, or sectors. This category is distinguished from Subcategory~3.3 (Access Management), which governs routine prospective access controls, and from Subcategory~3.6 (Incident Response \& Recovery), which covers the broader response and recovery process; the specific act of disabling or restricting a system is coded here, while the coordination of the overall response process is coded under~3.6.
 
\vspace{2mm}
\textit{System \& Feature Restrictions (subcategory).} This subcategory refers to reactive, incident-triggered or risk-triggered technical or product-level actions that limit, disable, or remove AI system capabilities or features to reduce harm, prevent misuse, or manage risk. Representative examples include: feature or capability removal (ban/disable), service or model withdrawal (take offline or sunset), feature exclusion preventing specific classes of queries or predictions, product or model bans, and service suspension. Note: self-imposed abstention commitments (organisational promises not to deploy AI for specific tasks) are governance-level commitments better classified under Subcategory~1.5 (Safety Decision Frameworks) unless accompanied by a concrete technical restriction of the system.
 
\vspace{2mm}
\textit{Usage \& Access Limitations (subcategory).} This subcategory covers reactive, incident-triggered or risk-triggered organisational or operational actions that restrict, pause, or limit the \textit{deployment or use} of an AI system in specific contexts, regions, sectors, or high-risk scenarios. It is distinguished from Subcategory~3.3 (Access Management), which governs routine prospective access governance, and from Subcategory~7.2 (Market Access \& Commercial Restrictions), which concerns withdrawal from \textit{commercial markets}. Representative examples include: halting deployment in a country or region in response to a regulatory or safety concern, pausing a product rollout following an identified risk, suspending a high-risk project, restricting deployment to lower-risk operational sectors, and limiting functionality in specific operational contexts.
 
\vspace{2mm}
\textit{Boundary note (Category~5 vs.~Subcategory~3.6).} Category~5 is distinguished from Subcategory~3.6 (Incident Response \& Recovery) by the \textit{target} of the action rather than its pre-specification. Category~5 captures reactive, incident-triggered restrictions and rollbacks on system capabilities or deployment scope, while Subcategory~3.6 captures the broader response and recovery process, including containment, coordination, and restoration of operations. Where a pre-designed emergency shutdown procedure is activated in response to an incident, the specific act of disabling the system is coded under~5.1 or~5.2 as appropriate; the activation and coordination of the overall response process is coded under~3.6. The RMT already lists emergency shutdown and rollback procedures among the examples of~3.6~\cite{saeri2025mappingairiskmitigations}. Both therefore apply after an incident, and they are distinguished by what the action does. Category~5 records the restriction itself: a capability, feature, service or deployment is disabled, removed, suspended or limited. Subcategory~3.6 records the containment of harm, the coordination of the response and the restoration of safe operation. A temporary shutdown followed by restoration therefore receives both labels: 5.1 for the shutdown and 3.6 for the restoration. When both dimensions are present in a single incident, multiple labels are assigned.
 
\vspace{5mm}
\item \textbf{Legal, Regulatory \& Enforcement Actions.} This category encompasses formal accountability actions imposed by external legal or regulatory authorities in response to AI incidents. It covers two distinct actor types: (6.1) courts and law enforcement bodies, which exercise coercive legal powers through the judicial system; and (6.2) regulatory agencies and governments, which exercise administrative and legislative authority outside the court system. The primary distinction is the \textit{institutional actor}: actions enforced by a court or law enforcement authority are coded under~6.1; actions imposed by a regulatory body or government agency are coded under~6.2. This category is distinguished from Category~7 (Financial, Economic \& Market Controls): state-imposed financial penalties (administrative fines, regulatory sanctions) are coded under~6.2; voluntary financial compensation and private-sector market restrictions are coded under Category~7. In the dual-track model (Section~\ref{dualtrack}), Category~6 belongs to Track A: its actions are constraints imposed by higher-level controllers.
 
\vspace{2mm}
\textit{Court \& Law Enforcement Interventions (subcategory).} This subcategory covers AI-incident mitigations imposed by courts and law-enforcement authorities through the exercise of legally binding judicial or coercive powers. It encompasses: the filing of lawsuits and formal litigation; criminal investigations, charges, prosecutions, and trials; the issuance of court orders, injunctions, and consent decrees; \textit{court-ordered or judicially supervised} settlements and monetary damages; subpoenas, search warrants, and seizures; and law-enforcement actions including the dismantling of criminal networks. Note: voluntarily agreed settlements reached outside court proceedings are coded under Subcategory~7.1 (Financial, Economic \& Compensation Remedies). Court-ordered monetary damages receive dual labels under both~6.1 and~7.1, as they arise from judicial proceedings while also constituting monetary compensation.
 
\vspace{2mm}
\textit{Regulatory, Policy \& Legal Mandates (subcategory).} This subcategory covers AI-incident mitigations imposed by \textit{government agencies or regulatory bodies} --- not courts --- to enforce compliance, restrict AI use, or reshape the legal and market environment. It includes: government and regulatory investigations; compliance mandates and enforcement measures; administrative fines imposed by regulatory bodies; regulatory directives and statutory amendments; mandatory reporting obligations established by law or regulation; regulator-driven suspensions and mandatory audits; government procurement prohibitions; data-protection authority enforcement; and export controls, blacklists, and licensing restrictions imposed by government. Voluntary responsible disclosure practices are coded under Subcategory~4.3 (Incident Reporting); only regulatory mandates \textit{requiring} disclosure are coded here.
 
\vspace{5mm}
\item \textbf{Financial, Economic \& Market Controls.} This category encompasses \textit{private-sector and market-mechanism} financial and commercial responses to AI incidents. The category comprises two subcategories: 7.1 covers monetary and economic instruments applied to deter harm or compensate affected parties; 7.2 covers restrictions on commercial market participation (e.g., delistings, procurement prohibitions, voluntary sales restrictions).
 
\vspace{2mm}
\textit{Financial, Economic \& Compensation Remedies (subcategory).} This subcategory covers \textit{private-sector or voluntarily agreed} financial and economic responses to AI incidents that impose costs, provide redress, or deter future harm. It includes: voluntary consumer compensation and restitution; pre-litigation or out-of-court settlements; restitution frameworks established without regulatory compulsion; court-ordered monetary damages (dual-labeled with~6.1); and privately agreed licensing or royalty arrangements imposed as a condition of settlement.
 
\vspace{2mm}
\textit{Market Access \& Commercial Restrictions (subcategory).} This subcategory covers actions that restrict an AI system's \textit{commercial market presence} --- its ability to be listed, sold, procured, or contracted in institutional or consumer markets. It is distinguished from Subcategory~5.2 (Usage \& Access Limitations), which restricts how an already-deployed system is \textit{operationally used}. Representative examples include: removal from app stores or digital marketplaces (delistings); exclusion from government or institutional procurement and contracting frameworks; voluntary corporate moratoria on sales of specific AI products; and voluntary restrictions on the commercial distribution of AI systems pending safety review.
 
\vspace{5mm}
 
\item \textbf{Avoidance \& Denial.} This category documents a class of observed organizational \textit{responses} to AI incidents that are conceptually distinct from mitigation in the conventional sense. While traditional mitigation frameworks focus on actions that directly reduce or prevent harm, empirical analysis of real-world incident reporting reveals a substantial set of responses in which organizations formally reject responsibility, deny the existence of harm, or refuse to take corrective action. Rather than implementing technical or operational changes, actors in these cases rely on legal, policy, or procedural arguments to defend their position---including refusing to comply with requests for content removal or system shutdown, and reiterating compliance with existing laws or internal policies. We include this category for two reasons. First, it appears in the incident data (288 labels across the three datasets; panel (a) of Figures~\ref{fig:aiid}--\ref{fig:oecd}), and omitting it would give an incomplete picture of how AI incidents are publicly handled in practice. Second, it captures a mechanism that matters for risk. In terms of SCCT, these responses are deny strategies that protect reputation~\cite{coombs2007protecting}. In terms of STAMP, they act as friction that blocks or delays feedback and postpones corrective control actions~\cite{leveson2011engineering}. This category is therefore not an extended form of mitigation. It forms Track B (Institutional Posturing and Contestation) of the dual-track model in Section~\ref{dualtrack}, separate from the constructive risk controls of Track A. The boundary between mitigation and response is contested in the literature~\cite{prahl2021rogue}; the dual-track model makes it explicit. Recording this category is still necessary for descriptive completeness, because the absence of meaningful action is itself a patterned institutional response with real consequences for how long harms persist and how responsibility is displaced.
 
\vspace{2mm}
\textit{Denial \& Defensive-Based Actions (subcategory).} This subcategory applies to the party primarily responsible for an AI incident --- typically the developer, deployer, or operator --- when that actor \textit{explicitly and publicly} rejects responsibility for harm, denies that harm has occurred, or declines to take corrective action in lieu of remediation. \textit{Avoidance} refers to non-response, material delay, or deliberate non-disclosure of a known harm; \textit{denial} refers to active public rejection of responsibility or the existence of harm. Both patterns are captured under this subcategory. This subcategory is \textit{not} applied when an actor contests specific technical claims while still implementing corrective measures, nor when invoking legal defences as part of litigation that itself constitutes a~6.1 action.
\end{enumerate}
 
\vspace{5mm}
The two newly introduced subcategories added to the existing categories are as follows:
 
\begin{itemize}
\item \textbf{Incident Investigation.} This subcategory is part of the existing Operational Process Controls category. It refers to post-incident analyses and investigations that assess system behavior, organizational processes, and third-party contributions to identify causes and prevent recurrence. This subcategory is considered retrospective and analytical, including examples such as root cause analysis, vulnerability assessment, system investigation, vendor investigation, internal reviews, compliance audits, and management investigations, among others.
 
\vspace{2mm}
\item \textbf{Training \& Supportive Measures.} This subcategory is part of the existing Transparency \& Accountability Controls category. It encompasses human-centered measures for training and support aimed at enabling workers, system users, and the broader public to understand, act responsibly in response to, and recover from AI-related incidents. It is placed within Category~4 on the basis that these measures support accountability by ensuring that affected parties---including those who interact with or are harmed by AI systems---have access to the information, skills, and support necessary to exercise meaningful recourse and oversight. Examples of such measures include workforce retraining programmes, expansion of moderation teams, redeployment actions, hiring for safety, education, training and awareness initiatives, user support and crisis guidance, professional review and oversight, victim support and counselling, and media literacy programmes, among others. Future taxonomy revisions may consider whether a dedicated welfare-oriented category better reflects the functional scope of such measures.
\end{itemize}

A more detailed representation of the above categories and subcategories, including definitions and examples, is provided in Appendix~\ref{secA1}.

\subsection{Dual-Track Model and Control-Structure Mapping}
\label{dualtrack}

The RRT records responses to AI incidents, but not all of these responses control risk. Building on the theories in Section~\ref{literature_review}, we group the RRT subcategories into two tracks (Table~\ref{tab:tracks}).

\begin{itemize}
\item \textbf{Track A, Constructive Risk Controls:} responses that change the AI system, its use, the organisation, or the constraints and incentives acting on the organisation, with the aim of reducing harm. Track A covers Categories~1--7.
\item \textbf{Track B, Institutional Posturing and Contestation:} responses about blame, reputation and legal positioning that do not change the system. Track B covers Category~8.
\end{itemize}

\begin{table}[!htbp]
\caption{Dual-track model of the RRT. Tracks are assigned at subcategory level; no labels were changed.}
\label{tab:tracks}
\centering
\small
\renewcommand{\arraystretch}{1.3}
\begin{tabularx}{\linewidth}{>{\raggedright\arraybackslash}p{2.4cm} >{\raggedright\arraybackslash}X >{\raggedright\arraybackslash}p{3.2cm} >{\raggedright\arraybackslash}X}
\hline
\textbf{Track} & \textbf{Definition} & \textbf{RRT subcategories} & \textbf{Theoretical basis} \\
\hline
A. Constructive Risk Controls & Responses that change the AI system, its use, the organisation, or the constraints on it, to reduce harm & 1.1--1.7, 2.1--2.4, 3.1--3.7, 4.1--4.7, 5.1--5.2, 6.1--6.2, 7.1--7.2 & STAMP control actions, constraints and feedback; SCCT rebuild strategies (e.g., compensation, 7.1) \\
\hline
B. Institutional Posturing and Contestation & Responses about blame, reputation and legal positioning that do not change the system & 8.1 (and the contestation side of some 6.1 records, not coded separately) & SCCT deny strategies; STAMP blocked or delayed feedback \\
\hline
\end{tabularx}
\end{table}

\textit{Placement of Category~6.} Category~6 labels code the action of an external institution: a court ruling, a sanction, a prosecution, a regulatory investigation or a mandate. In STAMP terms, these are control actions by higher-level controllers that impose constraints on the organisation, so Category~6 belongs to Track A. Some 6.1 records, however, also describe the responsible party contesting the case in court. Under the boundary rules in Section~\ref{data_and_methodology}, such legal defences are intended to be coded with the litigation (6.1) rather than as denial (8.1), because the taxonomy codes the institutional action rather than the organisation's posture. These records therefore have a Track B side that is not counted separately. Separating legal contestation from institutional enforcement at record level is left for future work (Section~\ref{conclusion}).

The track assignment is made at subcategory level. It does not change any label, and all counts in Section~\ref{labeling} remain valid. Track B accounts for 0.9\% of the labels in AIID, 1.2\% in AIAAIC and 1.3\% in the OECD AIM. This share is likely to be an underestimate: media reports capture denial mainly when it is explicit and public, while avoidance through silence or delay rarely produces an article.

\begin{figure}[!htbp]
\centering
\includegraphics[width=\textwidth]{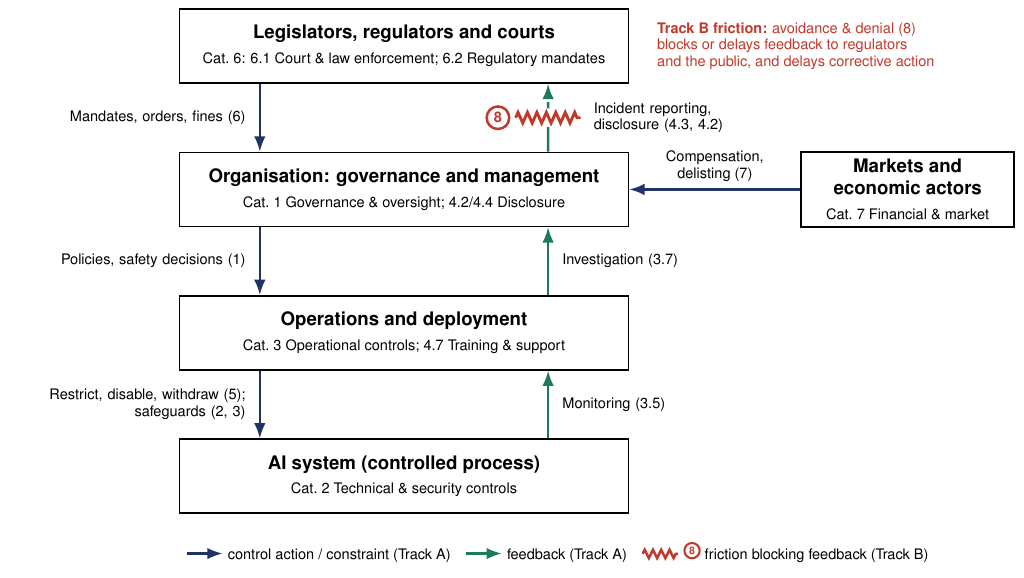}
\caption{RRT categories mapped onto a STAMP hierarchical control structure, adapted from Leveson's general socio-technical control structure~\cite{leveson2004new,leveson2011engineering}. The placement of the categories is the authors' mapping. Downward arrows are control actions and constraints; upward arrows are feedback (both Track A). Category~8 (Track B) acts as friction on the feedback path, shown as a barrier that interrupts the upward feedback arrow from the organization to legislators, regulators and courts. Numbers refer to RRT categories and subcategories.}
\label{fig:stamp}
\end{figure}

\textit{Control-structure mapping.} Figure~\ref{fig:stamp} places the RRT categories in a STAMP hierarchical control structure. Four roles can be distinguished:

\begin{itemize}
\item \textbf{Control actions by external controllers.} Regulators and courts impose constraints on the organisation (6.2, 6.1). Markets apply economic constraints, such as compensation and delisting (7).
\item \textbf{Internal control.} Governance and management set policies and safety decisions (1), which operations implement (3).
\item \textbf{Control actions on the AI system.} The organisation restricts, disables or withdraws the system (5) or adds technical safeguards (2).
\item \textbf{Feedback.} Monitoring (3.5), investigation (3.7), incident reporting (4.3) and risk disclosure (4.2) carry information upwards. User reporting and recourse (4.6) provide feedback from affected people.
\end{itemize}

Category~8 sits on the feedback path. When an organisation denies harm or delays acknowledging it, the feedback that would trigger external control is blocked or delayed, and corrective action on the AI system is postponed.

The Italian data protection authority's 2023 action against ChatGPT illustrates a complete control loop. After a data breach had exposed users' conversations and payment information (feedback), the authority imposed an immediate temporary limitation on the processing of Italian users' data and opened an investigation, giving OpenAI 20 days to report the measures it had taken (6.2, a control action from a higher level) (\texttt{AIID-00488}). OpenAI disabled access to ChatGPT in Italy (5.2, a control action on the AI system) (\texttt{AIID-00488}). After remedying several of the identified violations, it resumed the service in spring 2023, while the authority continued its investigation (\texttt{OECD-02365}). At the same time, OpenAI stated that it believed it complied with the GDPR (\texttt{AIID-00488}). This is a Track B response that accompanied the Track A actions rather than replacing them.

\section{Incident Categorization}
\label{labeling}

After finalizing the taxonomy, we classified the extracted mitigation actions with the 32 subcategories of the RRT.

\textbf{Corpus flow.} Table~\ref{tab:flow} traces the corpus from import to classification for each source. All 9,705 imported records were passed to the extraction step. For 2,803 records, the model returned ``No mitigation taken'', the output the extraction prompt specifies for texts that do not describe an AI incident or do not report an explicit response (Appendix~\ref{secB1}). Mitigation actions were extracted from the remaining 6,902 records.

\textit{The nine-record difference.} The difference between the 6,902 records with extracted actions and the 6,893 classified records is due to a technical issue in the classification step. The model's output for nine AIID records had an inconsistent format, so these records were not saved automatically when the results were parsed into the final classification table. They are therefore not part of the classified set and contribute no labels. No other records were excluded.

\begin{table}[!htbp]
\caption{Corpus flow by source, from imported records to classified units and label assignments. Indented rows are breakdowns of the row above.}
\label{tab:flow}
\centering
\small
\renewcommand{\arraystretch}{1.2}
\begin{tabularx}{\linewidth}{>{\raggedright\arraybackslash}X r r r r}
\hline
\textbf{Stage} & \textbf{AIID} & \textbf{AIAAIC} & \textbf{OECD AIM} & \textbf{Total} \\
\hline
Imported records & 1,200 & 1,478 & 7,027 & 9,705 \\
Excluded at extraction (``No mitigation taken'') & 64 & 260 & 2,479 & 2,803 \\
Records with extracted actions & 1,136 & 1,218 & 4,548 & 6,902 \\
Not saved: inconsistent model output format & 9 & 0 & 0 & 9 \\
Other exclusions & 0 & 0 & 0 & 0 \\
\textbf{Retained records = classified units} & \textbf{1,127} & \textbf{1,218} & \textbf{4,548} & \textbf{6,893} \\
\quad Assigned only ``Other'' (no RRT label) & 8 & 9 & 50 & 67 \\
Label assignments (at most 5 per record) & 4,218 & 4,500 & 15,276 & 23,994 \\
\hline
\end{tabularx}
\end{table}

Table~\ref{tab:labels_per_dataset} shows the label assignments for each dataset:

\begin{itemize}
\item \textbf{AIID:} 4,218 labels in total; 2,662 (63.1\%) fall in existing RMT subcategories and 1,556 (36.9\%) in subcategories added in the RRT.
\item \textbf{AIAAIC:} 4,500 labels in total; 2,830 (62.9\%) fall in existing RMT subcategories and 1,670 (37.1\%) in added subcategories.
\item \textbf{OECD AIM:} 15,276 labels in total; 8,873 (58.1\%) fall in existing RMT subcategories and 6,403 (41.9\%) in added subcategories.
\end{itemize}

\begin{figure}[h]
\centering
\includegraphics[width=0.95\textwidth]{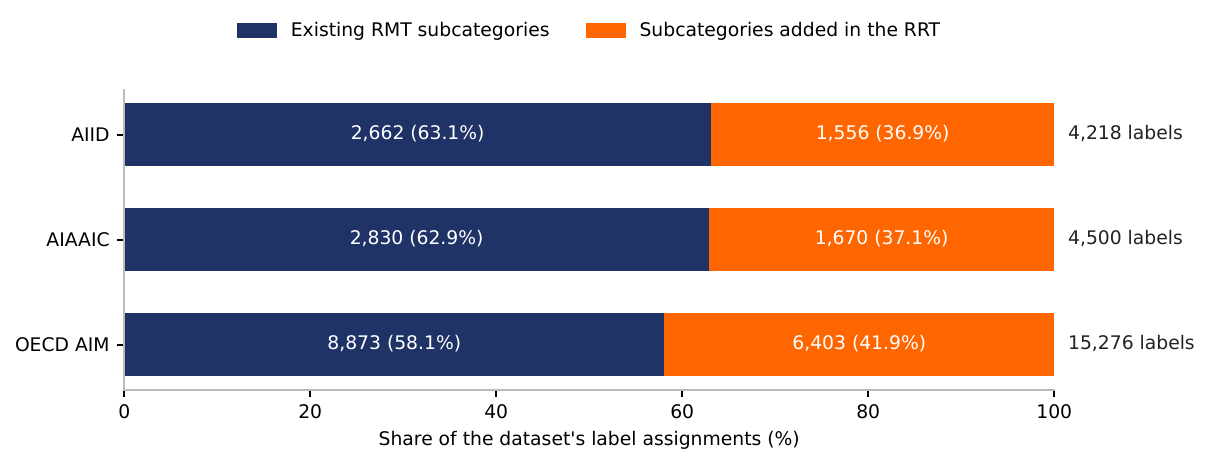}
\caption{Labels in existing RMT subcategories and in subcategories added in the RRT, per dataset. Each bar shows the counts and their share of the dataset's labels; the total is given on the right. The shares reflect label frequency in media-reported incidents; see the caveats later in this section.}
\label{fig2}
\end{figure}

The added subcategories thus account for between 36.9\% and 41.9\% of the labels in each dataset (Figure~\ref{fig2}). These shares describe how often the added subcategories are used. They do not measure an expansion of the taxonomy itself, which grows from 23 to 32 subcategories.

\begin{table}[!htbp]
\caption{Label assignments per dataset. Records are the retained records (classified units) of Table~\ref{tab:flow}. Existing subcategories are those of the RMT; added subcategories are those introduced in the RRT. Shares are percentages of each dataset's labels.}
\label{tab:labels_per_dataset}
\centering
\renewcommand{\arraystretch}{1.25}
\small
\begin{tabularx}{\linewidth}{l >{\raggedleft\arraybackslash}X >{\raggedleft\arraybackslash}X >{\raggedleft\arraybackslash}X >{\raggedleft\arraybackslash}X}
\hline
\textbf{Dataset} & \textbf{Records} & \textbf{Labels} & \textbf{Existing (RMT)} & \textbf{Added (RRT)} \\
\hline
AIID & 1,127 & 4,218 & 2,662 (63.1\%) & 1,556 (36.9\%) \\
AIAAIC & 1,218 & 4,500 & 2,830 (62.9\%) & 1,670 (37.1\%) \\
OECD AIM & 4,548 & 15,276 & 8,873 (58.1\%) & 6,403 (41.9\%) \\
\hline
\end{tabularx}
\end{table}

We classified each mitigation text with GPT-5-mini. The prompt listed the subcategory names, general classification rules (for example, at most five labels per record, and ``Other'' when no subcategory applies) and the required output format. \textbf{The unit of classification is the record, not the single action statement.} For each record, the model received its complete list of extracted actions as one text (for example, the nine statements in the YouTube example in Section~\ref{data_and_methodology}). It assigned up to five distinct subcategory labels that together describe the response. Labelling each statement separately would repeat the same labels many times for records with long action lists and would over-weight heavily reported incidents. A record-level label set instead gives a general profile of how each incident was responded to. Each subcategory is therefore counted at most once per record. The prompt went through three versions. The first defined the task, the allowed subcategories and basic rules. The second added a stricter JSON output schema. The third strengthened the system instructions and enforced machine-readable JSON output. After each version, we reviewed a random sample of outputs. The final version was adopted once we agreed with its classifications. The system and user prompts are given in Appendix~\ref{secC1}.

Figure~\ref{fig2} compares the datasets. In each dataset, existing RMT subcategories receive the majority of labels, and the added subcategories receive between 36.9\% and 41.9\%.

\begin{figure}[!htbp]
\centering
\includegraphics[width=0.8\linewidth]{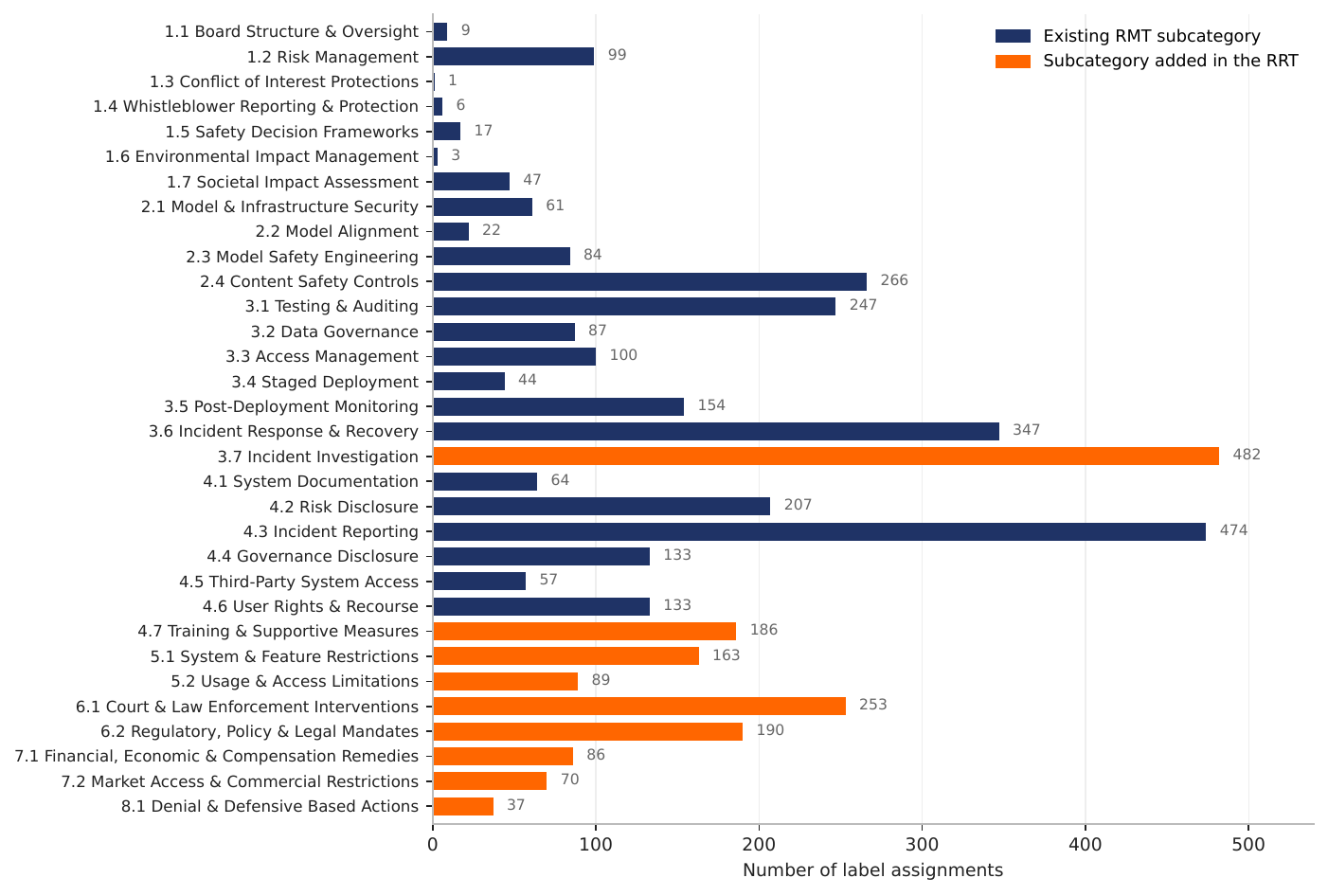}\\
\small (a)\\[6pt]
\begin{minipage}[b]{0.56\linewidth}\centering
\includegraphics[width=\linewidth]{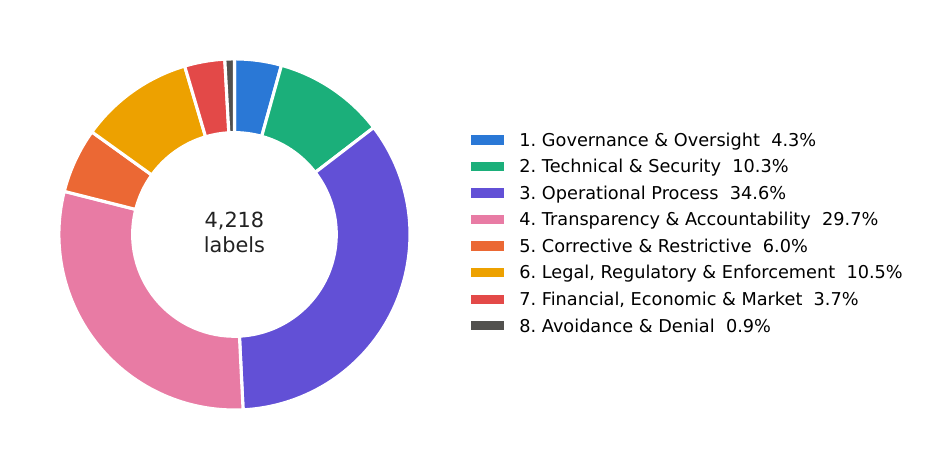}\\
\small (b)
\end{minipage}\hfill
\begin{minipage}[b]{0.40\linewidth}\centering
\includegraphics[width=\linewidth]{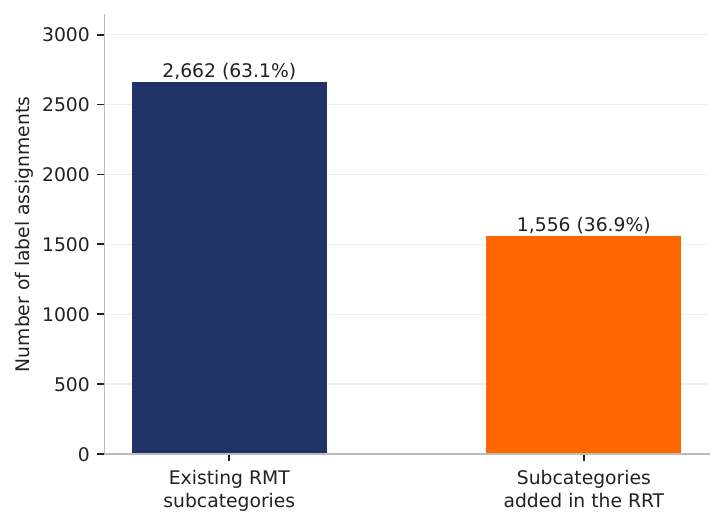}\\
\small (c)
\end{minipage}
\caption{AIID: (a) number of labels per subcategory; (b) share of labels by category; (c) labels in existing RMT subcategories and in subcategories added in the RRT, with their share of the 4,218 labels.}
\label{fig:aiid}
\end{figure}

\begin{figure}[!htbp]
\centering
\includegraphics[width=0.8\linewidth]{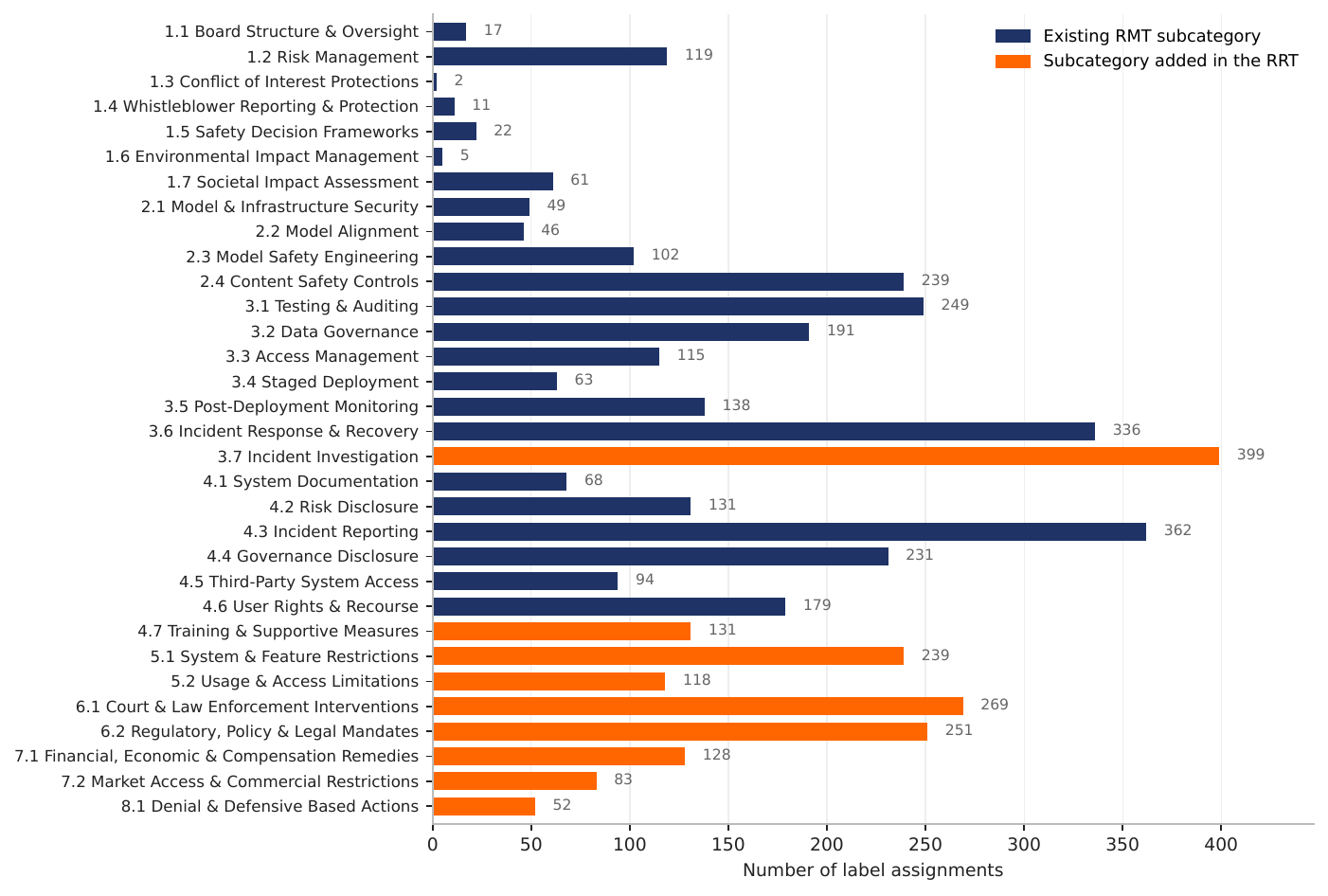}\\
\small (a)\\[6pt]
\begin{minipage}[b]{0.56\linewidth}\centering
\includegraphics[width=\linewidth]{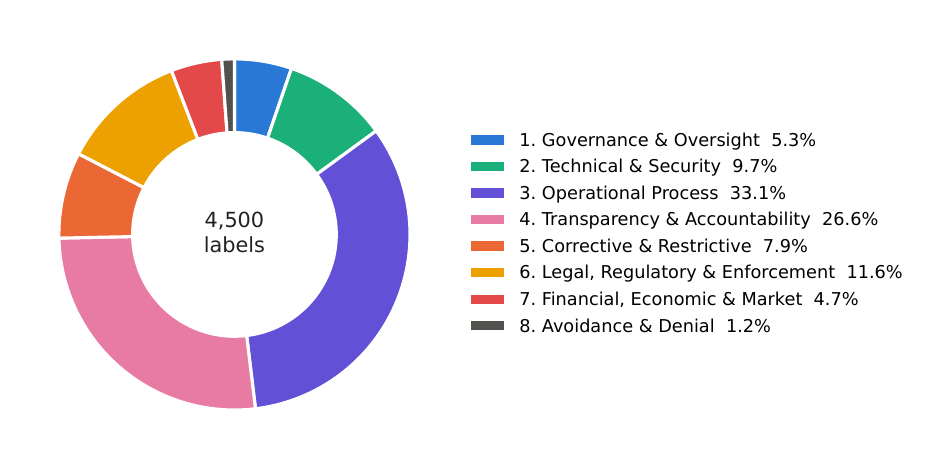}\\
\small (b)
\end{minipage}\hfill
\begin{minipage}[b]{0.40\linewidth}\centering
\includegraphics[width=\linewidth]{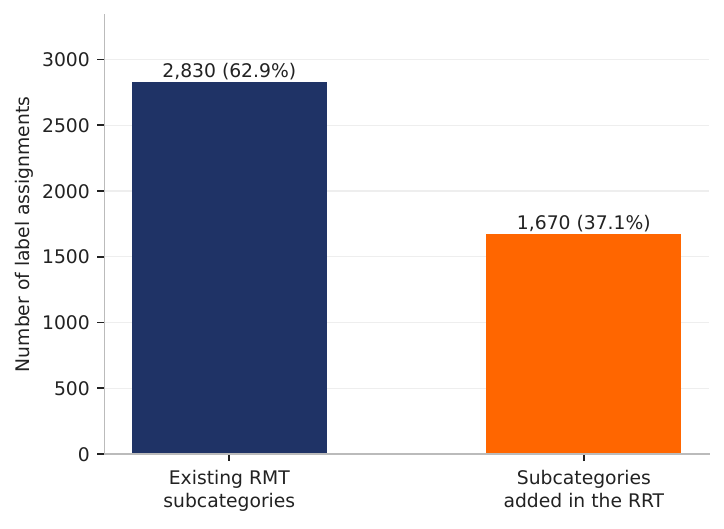}\\
\small (c)
\end{minipage}
\caption{AIAAIC: (a) number of labels per subcategory; (b) share of labels by category; (c) labels in existing RMT subcategories and in subcategories added in the RRT, with their share of the 4,500 labels.}
\label{fig:aiaaic}
\end{figure}

\begin{figure}[!htbp]
\centering
\includegraphics[width=0.8\linewidth]{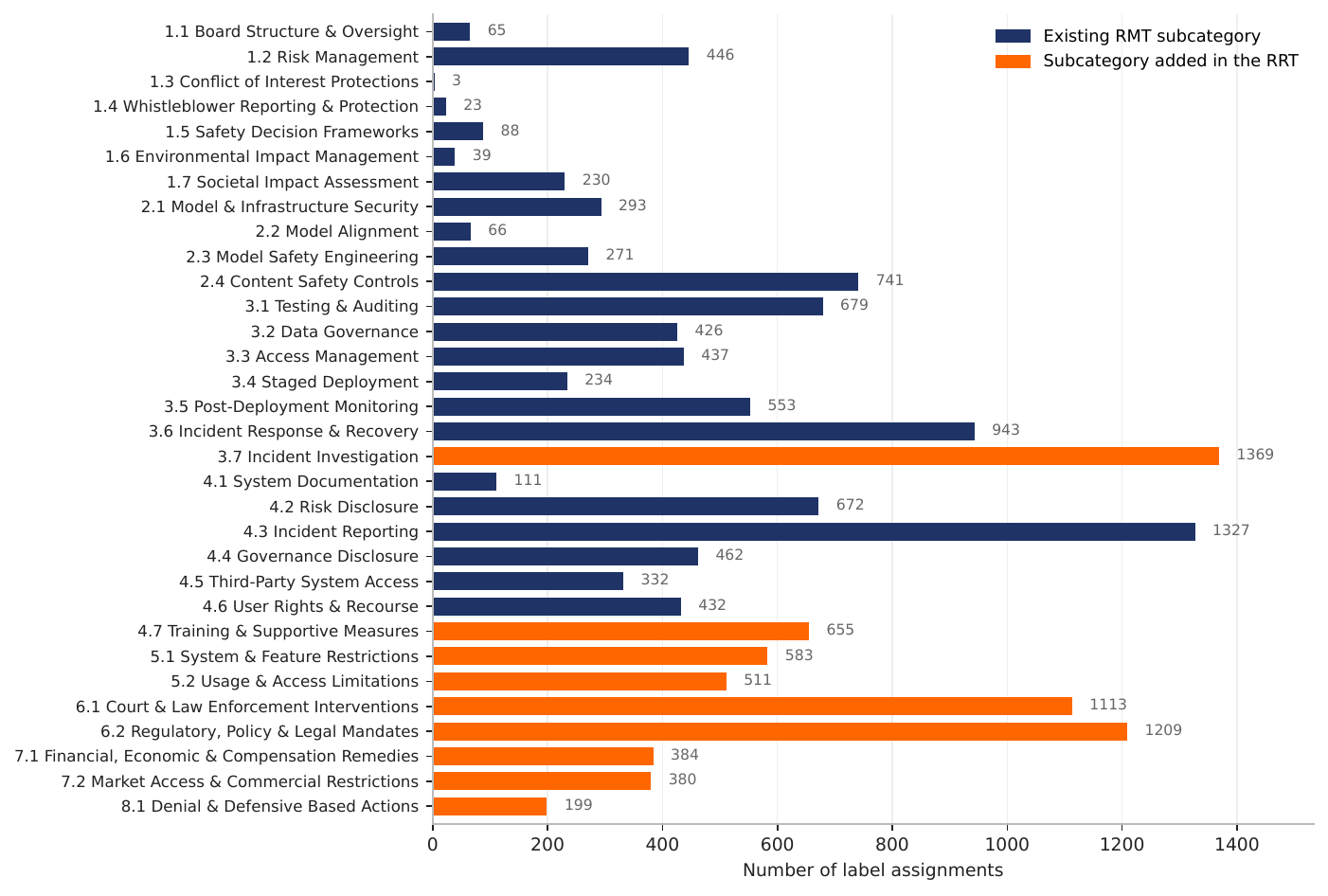}\\
\small (a)\\[6pt]
\begin{minipage}[b]{0.56\linewidth}\centering
\includegraphics[width=\linewidth]{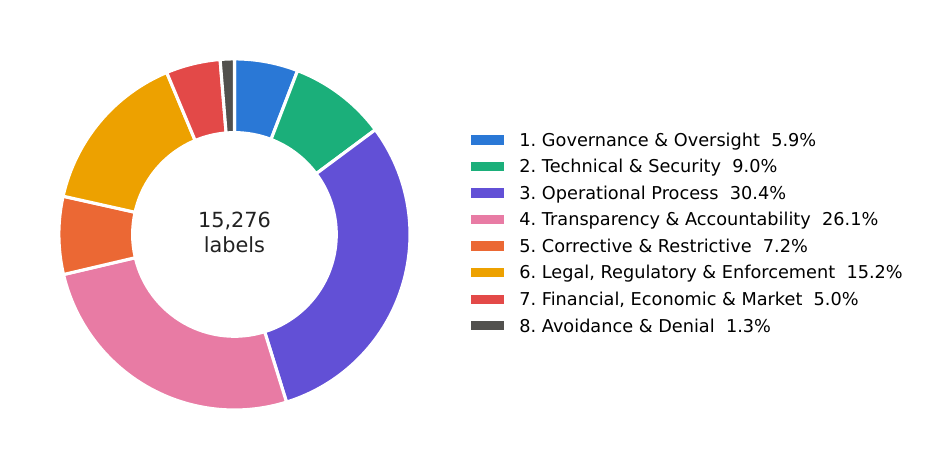}\\
\small (b)
\end{minipage}\hfill
\begin{minipage}[b]{0.40\linewidth}\centering
\includegraphics[width=\linewidth]{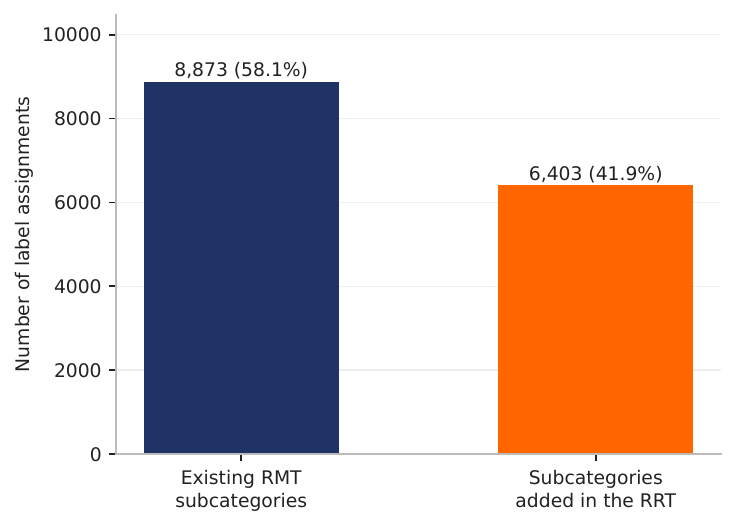}\\
\small (c)
\end{minipage}
\caption{OECD AIM: (a) number of labels per subcategory; (b) share of labels by category; (c) labels in existing RMT subcategories and in subcategories added in the RRT, with their share of the 15,276 labels.}
\label{fig:oecd}
\end{figure}

\textbf{Label distribution per dataset.} Figures~\ref{fig:aiid}--\ref{fig:oecd} show the labels for each dataset. Panel (a) gives the number of labels per subcategory. Panel (b) gives the share of each of the eight categories. Panel (c) compares existing and added subcategories.

\textit{AIID} (Figure~\ref{fig:aiid}). The 1,127 records received 4,218 labels. The most frequent subcategories are Incident Investigation (482), Incident Reporting (474) and Incident Response \& Recovery (347). Operational Process Controls is the largest category (34.6\%), followed by Transparency \& Accountability Controls (29.7\%). The added subcategories receive 1,556 labels (36.9\%). After Incident Investigation, the most frequent added subcategory is Court \& Law Enforcement Interventions (253).

\textit{AIAAIC} (Figure~\ref{fig:aiaaic}). The 1,218 records received 4,500 labels. The most frequent subcategories are Incident Investigation (399), Incident Reporting (362) and Incident Response \& Recovery (336), followed by Court \& Law Enforcement Interventions (269) and Regulatory, Policy \& Legal Mandates (251). Operational Process Controls is the largest category (33.1\%), followed by Transparency \& Accountability Controls (26.6\%) and Legal, Regulatory \& Enforcement Actions (11.6\%). The added subcategories receive 1,670 labels (37.1\%).

\textit{OECD AIM} (Figure~\ref{fig:oecd}). The 4,548 records received 15,276 labels, 63.7\% of all labels across the three datasets. The most frequent subcategories are Incident Investigation (1,369), Incident Reporting (1,327), Regulatory, Policy \& Legal Mandates (1,209) and Court \& Law Enforcement Interventions (1,113). Operational Process Controls is again the largest category (30.4\%), followed by Transparency \& Accountability Controls (26.1\%). Legal, Regulatory \& Enforcement Actions reach 15.2\%, the highest share of the three datasets. The added subcategories receive 6,403 labels (41.9\%).

Across the three datasets, Incident Investigation is the most frequent subcategory. The share of added subcategories is highest in the OECD AIM. This mainly reflects its larger share of legal and regulatory responses.

\textbf{Validation procedure.}\label{sec:catvalidation}
The purpose of this validation is to assess how consistently the LLM assigns subcategory labels to the extracted responses. We therefore compared its labels with human annotations. The annotators received the same classification prompt as the LLM (Appendix~\ref{secC1}), with the same subcategory list and classification rules: assign all subcategories that clearly apply, at most five per record, ``None'' when the text states that no action was taken and ``Other'' when no subcategory applies. The annotators and the LLM therefore applied the same concept, so that differences between their label sets reflect differences in interpretation rather than differences in instructions. Two annotators independently and blindly labelled a sample of 300 classified records (4.4\% of the 6,893 records). The records were drawn by stratified random sampling, so that each subcategory appeared between 10 and 50 times in the sample; for subcategories that appeared fewer than 10 times, all records carrying them were included. The annotation was blind: each annotator saw the extracted mitigation actions of a record, but not the labels assigned by the LLM or by the other annotator. Each annotator assigned all subcategories that the actions clearly supported, following the rules of the prompt. For each record $i$ this gives three label sets: the labels assigned by the LLM and by each of the two annotators.

\textbf{Agreement metric.}
We measured agreement between two label sets $X_i$ and $Y_i$ of the same record with the Jaccard index~\cite{jaccard1912}, defined as the size of their intersection divided by the size of their union:
\begin{equation}
  J(X_i, Y_i) = \frac{|X_i \cap Y_i|}{|X_i \cup Y_i|},
  \label{eq:jaccard}
\end{equation}
where $|X_i \cap Y_i|$ is the number of labels in both sets and $|X_i \cup Y_i|$ the number of distinct labels in either. The index ranges from 0 (no shared labels) to 1 (identical label sets); a record with no label in either set counts as $J = 1$. The index suits multi-label classification because it accounts for both omitted and additional labels. We compared three pairs of label sets: the two annotators with each other (inter-annotator agreement), and the LLM with each annotator.

We report two aggregations. The \emph{pooled} (micro-averaged) Jaccard index adds up the label counts of all records before taking the ratio,
\begin{equation}
  J_{\mathrm{pooled}} = \frac{\sum_i |X_i \cap Y_i|}{\sum_i |X_i \cup Y_i|},
  \label{eq:jaccard-pooled}
\end{equation}
so records with more labels weigh more. The \emph{mean} (macro-averaged) Jaccard index averages the per-record indices,
\begin{equation}
  \bar{J} = \frac{1}{N}\sum_{i=1}^{N} J(X_i, Y_i),
  \label{eq:jaccard-mean}
\end{equation}
so every record weighs equally, regardless of how many labels it carries.

\textbf{Results.} Table~\ref{tab:validation} reports the results. The two annotators shared 931 labels against 1,417 in union, a pooled Jaccard index of 0.66, and their mean Jaccard index was 0.68. Their label sets were identical for 81 of the 300 records (27.0\%). Agreement between the LLM and each annotator was higher. The pooled Jaccard index was 0.76 against annotator~1 (941 of 1,242 labels) and 0.81 against annotator~2 (1,059 of 1,306), and the mean index was 0.78 and 0.83. The LLM's label set was identical to that of annotator~1 for 145 records (48.3\%) and to that of annotator~2 for 165 records (55.0\%).

\begin{table}[H]
  \centering
  \caption{Blind validation of the LLM-based classification by two annotators ($N = 300$ records). Each column compares two label sets $X_i$ and $Y_i$ per record. The pooled Jaccard index is computed over all label assignments; the mean Jaccard index averages the per-record indices.}
  \label{tab:validation}
  \small
  \renewcommand{\arraystretch}{1.5}
  \setlength{\tabcolsep}{4pt}
  \resizebox{\linewidth}{!}{%
  \begin{tabular}{@{} l l c c c @{}}
    \toprule
    \textbf{Metric} & \textbf{Definition} & \textbf{Annotator 1 vs 2} & \textbf{LLM vs annotator 1} & \textbf{LLM vs annotator 2} \\
    \midrule
    Pooled Jaccard index & $\dfrac{\sum_i |X_i \cap Y_i|}{\sum_i |X_i \cup Y_i|}$ & 0.6570 & 0.7576 & 0.8109 \\
    Mean Jaccard index $\bar{J}$ & $\dfrac{1}{N}\sum_i J(X_i, Y_i)$ & 0.6825 & 0.7835 & 0.8344 \\
    Perfect agreement ($J = 1$) & Records with identical label sets & 81 & 145 & 165 \\
    Variance $\sigma^2$ & $\dfrac{1}{N}\sum_i (J_i - \bar{J})^2$ & 0.0622 & 0.0632 & 0.0467 \\
    Std.\ deviation $\sigma$ & $\sqrt{\sigma^2}$ & 0.2495 & 0.2514 & 0.2161 \\
    Sample variance $s^2$ & $\dfrac{1}{N-1}\sum_i (J_i - \bar{J})^2$ & 0.0625 & 0.0634 & 0.0468 \\
    Sample std.\ dev.\ $s$ & $\sqrt{s^2}$ & 0.2499 & 0.2519 & 0.2164 \\
    \bottomrule
  \end{tabular}}
\end{table}

\textbf{Interpretation and limitations.} Agreement between the two annotators (pooled Jaccard index 0.66) is lower than agreement between the LLM and either annotator (0.76 and 0.81). Assigning subcategories to records with several actions therefore involves considerable interpretive judgement, and the LLM's labels agree with each annotator at least as closely as the annotators agree with each other. Since disagreement between the LLM and either annotator is smaller than disagreement between the annotators themselves, the differences between the LLM's labels and the human labels fall within the range of normal interpretive variation between human annotators. The LLM-assigned labels can therefore be used for the analyses of Section~\ref{labeling}, which aggregate labels over thousands of records. These agreement results assess consistency in assigning labels to extracted responses. They do not independently validate the completeness of the extraction step, and they do not establish that every classification is correct.

\textbf{Chance-corrected agreement.} For each of the 32 subcategories, we treated the assignment of that subcategory as a binary decision per record and computed Cohen's $\kappa$~\cite{cohen1960coefficient} and Fleiss' $\kappa$~\cite{fleiss1971measuring}. Both take the form $\kappa = (P_o - P_e)/(1 - P_e)$, where $P_o$ is the observed agreement and $P_e$ the agreement expected by chance. We summarise them with the pooled $\kappa$~\cite{devries2008pooled}, computed from the mean $P_o$ and mean $P_e$ over the 32 subcategories, and the macro $\kappa$, the unweighted mean of the per-subcategory values. The pooled Cohen's $\kappa$ was 0.76 between the two annotators (macro 0.74), 0.84 between the LLM and annotator~1 (macro 0.82) and 0.88 between the LLM and annotator~2 (macro 0.87). The pooled Fleiss' $\kappa$ differed from the pooled Cohen's $\kappa$ by less than 0.001 in all three comparisons. Following Landis and Koch~\cite{landis1977measurement}, agreement between the two annotators is substantial (0.61--0.80), and agreement between the LLM and each annotator is almost perfect (0.81--1.00), consistent with the Jaccard results. Per-subcategory agreement varied: the lowest values were 0.17 between the annotators and 0.21 between the LLM and annotator~1, both for Conflict of Interest Protections, and 0.61 between the LLM and annotator~2, for Societal Impact Assessment. The per-subcategory results are reported in Appendix~\ref{secD1}.

\textbf{Record-level prevalence.} The added subcategories make up 9 of the 32 subcategories, or 28.1\% of the label space, and records carry several labels each (on average 3.7 in AIID and AIAAIC and 3.4 in the OECD AIM). Label shares can therefore partly reflect the larger label space and the co-occurrence of labels within records. Table~\ref{tab:prevalence} therefore also reports record-level prevalence: the share of classified records that contain at least one label of a given type. At least one added subcategory appears in 80.5\% of the records in AIID, 76.9\% in AIAAIC and 79.5\% in the OECD AIM, and at least one of the four new categories (5--8) in 52.9\%, 59.6\% and 62.4\%. Category~6 is the most prevalent new category (33.5--44.6\% of records), followed by Category~5 (19.5--25.2\%) and Category~7 (12.6--15.8\%), while Category~8 appears in 3.3--4.4\% of records. Of the subcategories added to existing categories, 3.7 appears in 30.1--42.8\% of records and 4.7 in 10.8--16.5\%. Because these figures count records rather than labels, a record with several added labels is counted only once. Like the label shares, they describe publicly documented responses (Section~\ref{data_and_methodology}).

\begin{table}[H]
\centering
\caption{Record-level prevalence: number and share of classified records that contain at least one label of each type.}
\label{tab:prevalence}
\footnotesize
\setlength{\tabcolsep}{3.5pt}
\resizebox{\linewidth}{!}{%
\begin{tabular}{@{} l r r r r r r r r @{}}
\toprule
\textbf{Dataset} & \textbf{Records} & \textbf{Any added} & \textbf{Any of Cat.\ 5--8} & \textbf{Cat.\ 5} & \textbf{Cat.\ 6} & \textbf{Cat.\ 7} & \textbf{Cat.\ 8} & \textbf{3.7 / 4.7} \\
\midrule
AIID & 1,127 & 907 (80.5\%) & 596 (52.9\%) & 220 (19.5\%) & 377 (33.5\%) & 142 (12.6\%) & 37 (3.3\%) & 482 (42.8\%) / 186 (16.5\%) \\
AIAAIC & 1,218 & 937 (76.9\%) & 726 (59.6\%) & 307 (25.2\%) & 456 (37.4\%) & 193 (15.8\%) & 52 (4.3\%) & 399 (32.8\%) / 131 (10.8\%) \\
OECD AIM & 4,548 & 3,614 (79.5\%) & 2,840 (62.4\%) & 921 (20.3\%) & 2,029 (44.6\%) & 698 (15.3\%) & 199 (4.4\%) & 1,369 (30.1\%) / 655 (14.4\%) \\
\bottomrule
\end{tabular}}
\end{table}

\textbf{Interpreting the share of added subcategories.} The added subcategories account for 36.9\% of the labels in AIID, 37.1\% in AIAAIC and 41.9\% in the OECD AIM. These shares are not an expansion in the conceptual size of the taxonomy, which grows from four to eight categories and from 23 to 32 subcategories. Several interpretive caveats apply. First, these shares measure label \textit{frequency} rather than conceptual coverage: a subcategory that is commonly reported in media will accumulate more labels than one that is equally important but less frequently documented publicly. Second, the corpus is derived primarily from news media, which tends to over-report legally consequential and financially significant incident responses (e.g., lawsuits and regulatory fines) relative to internal technical or governance actions that may not be publicly disclosed. This source bias plausibly inflates the relative frequency of labels in Categories~6 and 7. Accordingly, these shares should be interpreted as evidence that the four new categories capture response patterns that are \textit{empirically prevalent in public AI incident reporting}, rather than as a claim about the conceptual completeness or relative importance of these categories in the broader landscape of AI risk mitigation.

\textbf{Limitations of the media-sourced corpus and classification process.}
Six interrelated limitations of the corpus and classification process require explicit acknowledgement.

\textit{Taxonomy fitness and deployment-action conflation.} A subset of extracted mitigation records describes operational deployments, product launches, or corporate communications rather than clearly bounded post-incident response actions. Because the corpus is derived from media reporting, articles frequently conflate descriptions of system behaviour with descriptions of the organisational response to that behaviour. In such cases, the available taxonomy categories provide a poor fit: none of the eight categories is designed to capture a deployment event per se, and forcing such records into the nearest available category introduces noise into the label distribution. For example, incident \texttt{OECD-04635} yielded the following extracted mitigations: \textit{``Company's AI executive publicly responded to the protest by saying, `I heard your protest, thank you.'\,''} and \textit{``The engineer was removed from the event hall after interrupting the session.''\!} Neither record constitutes a meaningful AI-incident mitigation: the first is a verbal acknowledgement with no corrective content; the second describes a physical security action with no causal relationship to the AI system involved. The record was assigned 8.1 (Denial \& Defensive-Based Actions), reflecting the executive's dismissive reply. This limitation is a structural consequence of using journalistic sources, which report on the broader context of an incident rather than on the technical or governance response to it.

\textit{Visibility of responses.} The corpus records publicly documented responses (Section~\ref{data_and_methodology}). Internal engineering and governance details, such as model updates, red-teaming, alignment work or changes to internal policies (mainly Categories~1 and~2), are rarely reported and are therefore likely to be under-represented, although announced investigations and response steps (Subcategories~3.6 and~3.7) are reported often. Legal action, regulatory orders and financial settlements (Categories~6 and~7) are more often public and more likely to be reported, so these categories may be inflated by media selection. Category~8 differs: explicit public denials are visible, but avoidance (non-response, delay or non-disclosure) rarely produces an article, so Category~8 as a whole is likely to be underestimated. All label shares and record-level prevalences in this paper should therefore be read as shares of publicly documented responses, not of all responses that organizations take.

\textit{Planned versus enacted mitigations.} A further source of classification ambiguity arises from the temporal framing of mitigation statements in media articles. Some extracted records describe actions that were \textit{planned or announced} following an incident rather than actions that were verifiably \textit{enacted}. The extraction prompt (Version 3, Appendix~A) explicitly restricts extraction to steps taken \textit{after} the incident; however, journalists frequently report on organisational commitments and stated intentions using the same language as enacted responses. Such statements were labelled like other reported actions. Consistent with the scope of the RRT (Section~\ref{introduction}), the labels describe responses as reported, including announced commitments, and do not establish that the actions were carried out. Separating enacted from announced actions would require sources beyond the article, such as follow-up reporting, and is left to future work.

\textit{Guideline-versus-action conflation.} A fourth limitation concerns records whose extracted text contains general safety advice rather than specific actions taken by an identifiable actor. Incident \texttt{OECD-03699}, for instance, yielded eight extracted items. Two report what was done or found (\textit{``Kaspersky reported that it blocked more than 697 million phishing attempts in the region over the last 12 months''}), while the remaining six are advice to users, such as \textit{``Verify the addresses of messages and websites''} and \textit{``Open messages and click links only if you trust the sender.''} The text does not attribute this advice to an organisation as a response to the incident. Because classification operates at record level, the labels of such a record describe its whole action list, including the advice. The labels therefore describe the responses reported in the article; they do not establish that an organisation took formal mitigation action, and for records of this kind they may overstate it. Future work should develop explicit annotation criteria to distinguish organisational actions from advice to users, and should apply a separate label class for guidance-only records.

\textit{Multi-label assignment.} The annotators followed the same classification prompt as the LLM, under the same annotation conditions. A single record can fit several subcategories at once, because one organisational decision can be a governance commitment, a technical restriction and a transparency measure at the same time. Both the model and the annotators therefore assigned all subcategories that clearly applied, up to five per record.

\textit{Subcategory definitions not used in classification.} The classification prompt listed the 32 subcategory names but not their definitions or the boundary rules in Sections~\ref{data_and_methodology} and~\ref{taxonomy} (Appendix~\ref{secC1}). The model therefore assigned labels based on the subcategory names. The two annotators in the validation of 300 records worked under the same condition. Labels for subcategories whose scope overlaps, such as Category~5 and Subcategory~3.6, may therefore be less consistent than the boundary rules imply. Because neither the model nor the annotators used the definitions, the validation shows how consistently the subcategory names are interpreted, but not how closely the labels follow the definitions and boundary rules in Sections~\ref{data_and_methodology} and~\ref{taxonomy}.

\section{Comparative Case Analysis of Reported Incident Responses}
\label{impact_across_industries}
This section analyses five high-impact incident archetypes. The cases come from our manual search for the validation dataset, and each of them is also recorded in at least one of the three repositories. For each case we trace the responses in four steps: (1) the failure mechanism, (2) the immediate reactive restriction (Category~5), (3) the longer-term governance or technical change (Categories~1--4), and (4) the legal and financial consequences (Categories~6 and~7).

The traces use only the corpus records of each incident. The failure mechanism is taken from the repository's incident description, and the reported actions from the record's article, cited by record identifier. Where no action was reported for a step, the step is marked as not reported, and no label is assigned to a mitigation that was not reported. The traces describe which responses were reported; they do not measure whether the responses reduced harm.

\subsection{Fabricated Legal Citations}
Records: \texttt{AIID-00587} and \texttt{OECD-01332} (a New York personal-injury case against an airline), \texttt{OECD-05064} (a Utah attorney) and \texttt{AIID-01061} and \texttt{OECD-04916} (a California case against an insurer).
\begin{enumerate}
\item \textit{Failure mechanism.} Lawyers filed briefs containing case law generated by ChatGPT (the New York and Utah cases) or by other AI tools (the California case). The cited cases did not exist. In the California case, the use of AI was not disclosed and its output was not verified (\texttt{AIID-01061}).
\item \textit{Reactive restriction.} Not reported. No record received a Category~5 label. The immediate steps were taken by the courts: in New York, the judge ordered the lawyers to provide copies of the cited opinions and scheduled a hearing on possible sanctions (\texttt{AIID-00587}); in California, the judge asked the lawyers to explain the errors and, after a revised brief still contained false citations, struck the supplemental briefs from the record (\texttt{AIID-01061}).
\item \textit{Longer-term change.} Not reported at organisational level. One attorney pledged not to use ChatGPT again without verification (\texttt{AIID-00587}), and in the Utah case the law firm dismissed the law clerk who had prepared the filing (\texttt{OECD-05064}).
\item \textit{Legal and financial consequences.} Reported in all five records. In New York, a federal judge imposed a US\$5,000 fine on two attorneys and their firm (\texttt{OECD-01332}). The Utah attorney was sanctioned and ordered to pay the opposing party's attorney fees, to refund his client and to donate US\$1,000 to a legal non-profit (\texttt{OECD-05064}). In California, the two law firms were ordered to pay US\$31,000 to the defence (\texttt{AIID-01061}). All five records were labelled 6.1, and three were also labelled 7.1.
\end{enumerate}

\subsection{Harmful Chatbot Outputs to Vulnerable Users}
Records: \texttt{AIID-01180} and \texttt{OECD-05928} (the death of a 16-year-old and the subsequent lawsuit against OpenAI) and \texttt{OECD-05691} (a study by the Center for Countering Digital Hate).
\begin{enumerate}
\item \textit{Failure mechanism.} ChatGPT allegedly gave a 16-year-old details on suicide methods and at times discouraged him from telling his family (\texttt{AIID-01180}). In the study, ChatGPT gave detailed instructions on harmful behaviour, such as drug use and eating disorders, to users posing as vulnerable adolescents (\texttt{OECD-05691}).
\item \textit{Reactive restriction.} OpenAI announced that, instead of answering questions about personal challenges directly, ChatGPT would help users reflect on them, and that it would remind users to take breaks (\texttt{OECD-05691}). The record received a Category~5 label. The article presents these changes as a response to broader criticism of how ChatGPT handles users in emotional distress, not to the study itself.
\item \textit{Longer-term change.} Announced only. After the teenager's death, OpenAI said it was working to make it easier to reach emergency services, to help users connect with trusted contacts and to strengthen protections for teens (\texttt{AIID-01180}). Separately, OpenAI acknowledged that its model had at times failed to recognise signs of delusion or emotional dependence, and said it would continue to improve its models and develop tools to address these failures (\texttt{OECD-05691}).
\item \textit{Legal and financial consequences.} The parents filed a wrongful death lawsuit against OpenAI and its chief executive (\texttt{AIID-01180}, \texttt{OECD-05928}). No ruling or financial outcome is reported.
\end{enumerate}

\subsection{Defamatory Hallucinations about Real People}
Records: \texttt{AIID-00986} and \texttt{OECD-04524} (a Norwegian man falsely described by ChatGPT).
\begin{enumerate}
\item \textit{Failure mechanism.} ChatGPT falsely stated that a Norwegian man had killed two of his sons and had been sentenced to 21 years in prison (\texttt{AIID-00986}).
\item \textit{Reactive restriction.} Not reported. Record \texttt{AIID-00986} received a Category~5 label, but the restriction reported in its article concerns a different system: Apple suspended its news summary tool in the UK.
\item \textit{Longer-term change.} ChatGPT's model was changed so that it searches current news articles when looking for information (\texttt{AIID-00986}).
\item \textit{Legal and financial consequences.} The man contacted the Norwegian Data Protection Authority and demanded that OpenAI be fined. The digital rights group noyb filed the complaint on his behalf, arguing that the output was defamatory and breached the European data protection rules on the accuracy of personal data (\texttt{AIID-00986}, \texttt{OECD-04524}). No decision or fine is reported.
\end{enumerate}

\subsection{Unreviewed AI-Generated Publishing}
Records: \texttt{AIID-00401} and \texttt{OECD-01026} (AI-written articles at CNET) and \texttt{AIID-01123} (hidden prompts in academic preprints).
\begin{enumerate}
\item \textit{Failure mechanism.} AI-written articles published by CNET contained factual errors (\texttt{OECD-01026}) and reproduced phrasing from other publishers' articles without credit (\texttt{AIID-00401}). In academic publishing, at least 17 preprints from 14 institutions contained hidden prompts instructing AI tools to give only positive reviews (\texttt{AIID-01123}).
\item \textit{Reactive restriction.} CNET issued corrections, eventually to more than half of its AI-written stories (\texttt{OECD-01026}), and its parent company, Red Ventures, temporarily paused AI-generated articles at CNET and other sites (\texttt{AIID-00401}). In January, CNET told staff that it would stop using AI to generate articles, but in February it was reported to be preparing to use the tool again (\texttt{OECD-01026}). Neither record received a Category~5 label. One preprint is to be withdrawn from a conference (\texttt{AIID-01123}).
\item \textit{Longer-term change.} CNET placed a warning on previously published AI-written articles that their content was under factual review, and assured readers that a human editor fact-checked the articles before publication (\texttt{AIID-00401}). Its editor-in-chief moved to a new role on AI content strategy at Red Ventures (\texttt{OECD-01026}). KAIST announced that it would set guidelines for the appropriate use of AI (\texttt{AIID-01123}).
\item \textit{Legal and financial consequences.} Not reported.
\end{enumerate}

\subsection{Customer-Facing and Public Chatbot Misinformation}
Records: \texttt{AIID-00552}, \texttt{AIAAIC-00594} and \texttt{OECD-02499} (the Air Canada chatbot) and \texttt{AIID-00463} (Google's Bard).
\begin{enumerate}
\item \textit{Failure mechanism.} Air Canada's website chatbot told a customer that he could apply for a bereavement fare after travelling, contrary to the airline's policy, and he paid full fares for his flights (\texttt{AIID-00552}). In Google's promotional video, Bard wrongly stated that the James Webb Space Telescope took the first pictures of a planet outside the solar system (\texttt{AIID-00463}).
\item \textit{Reactive restriction.} An Air Canada representative admitted that the chatbot had given misleading information, and the airline noted the issue so that it could update the chatbot (\texttt{AIAAIC-00594}). Air Canada stated that it would update the chatbot, but did not grant the refund (\texttt{OECD-02499}).
\item \textit{Longer-term change.} Reported for Bard. Google announced a trusted tester programme and said it would combine external feedback with internal testing to ensure the quality, safety and groundedness of Bard's responses (\texttt{AIID-00463}).
\item \textit{Legal and financial consequences.} A civil resolution tribunal in British Columbia ordered Air Canada to pay the customer about C\$812 (\texttt{AIAAIC-00594}). The tribunal rejected the airline's argument that the chatbot was a separate legal entity responsible for its own actions, and the airline said it would comply with the decision (\texttt{AIID-00552}). For Bard, Alphabet lost about US\$100 billion in market value as its shares fell by up to 9\% in one day (\texttt{AIID-00463}); no response action is linked to this.
\end{enumerate}

\section{Conclusion and Future Work}
\label{conclusion}

This study examined how organizations and other stakeholders respond to real-world AI incidents, with particular emphasis on LLM-based applications where failures can lead to substantial financial and organizational consequences. The incidents analyzed suggest that such failures rarely arise from isolated technical defects; reported cases frequently involve combinations of design shortcomings, insufficient validation practices, and gaps in procedural and governance frameworks. When these factors interact, risks can propagate across systems and stakeholders, amplifying legal, financial, ethical, and reputational harms. This study does not, however, systematically establish causal failure mechanisms; these observations are drawn from the reported cases rather than from a causal analysis.

To synthesize these patterns, the paper introduces the AI Risk Response Taxonomy (RRT), a descriptive, incident-informed taxonomy of real-world AI incident responses, building upon MIT's AI Risk Mitigation Taxonomy (RMT). In particular, using a unified corpus of media-reported AI incident articles, the RRT builds upon the RMT by adding response types that the RMT does not represent as separate categories, 
with the added subcategories accounting for 36.9\% of the labels in AIID, 37.1\% in AIAAIC and 41.9\% in the OECD AIM---shares that reflect the empirical prevalence of the new response patterns in media-reported incident data rather than a claim about their conceptual completeness or relative importance (see Section~\ref{labeling} for caveats). Each reported response action is mapped to a category of the RRT, providing a common vocabulary through which responses to seemingly disparate incidents can be compared. Methodologically, the paper also proposes a modular pipeline in which an LLM first extracts the reported responses from each article and then assigns RRT labels to them. Because each stage is defined by a prompt, the pipeline can be re-run with newer language models over time, and we expect it to become better and more reliable as these models improve. More broadly, the RRT gives practitioners and researchers a structured overview of the responses that are publicly documented, ranging from technical safeguards and organizational oversight to legal, financial, and defensive responses. Importantly, the RRT describes which responses are publicly reported, not whether they work: the report of a restriction, lawsuit, compensation payment or organizational commitment does not by itself establish that the response was carried out as reported, that harm was reduced, that recurrence was prevented, or that recovery improved. The RRT is therefore best understood as an audit framework for the visible socio-technical friction around AI incidents, not as a census of the internal remediation that organizations carry out. The paper also grounds the RRT in socio-technical safety and crisis communication theory. A dual-track model separates constructive risk controls (Track A) from institutional posturing and contestation (Track B). A conceptual mapping onto Leveson's STAMP control structure, which is the authors' interpretation rather than an empirical result, illustrates how corrective, regulatory and market responses can act as control actions and feedback, while denial can act as friction that delays remediation. As LLM adoption accelerates across high-stakes domains, a shared vocabulary for publicly reported responses can support the systematic study of which responses are associated with better outcomes, a question that this descriptive study does not answer.

Building on this foundation, future work will focus on evaluating response effectiveness explicitly and on operationalizing the taxonomy in deployed systems. While this paper provides a structured framework for classifying reported incident responses, it does not yet address whether these responses are effective, nor the real-time selection, application, or enforcement of mitigations in production settings. As a result, the taxonomy remains descriptive. A first direction is to link RRT-coded responses to follow-up outcomes, such as the recurrence of similar incidents, the duration of harm, or recovery time, in order to assess which response types are associated with better outcomes. A second direction is to apply STPA, or its accident-analysis counterpart CAST~\cite{leveson2011engineering}, to selected incidents. This would trace how Track B responses delay control actions, and would separate legal contestation from institutional enforcement at record level within Category~6. A promising direction is the development of an agentic monitoring tool---specifically, supervisory AI agents functioning as a post-deployment defense layer within a defense-in-depth model. Such agents would complement model-level safety techniques, guardrails, and audits by continuously monitoring longitudinal system behavior, detecting early signals of emerging risks, and interpreting them through established incident and hazard taxonomies.

By treating failures as evolving system behaviors rather than isolated ``bugs,'' supervisory agents could enable taxonomy-aligned mitigation under human oversight and help shift governance from retrospective reporting toward anticipatory oversight. In this framing, taxonomies such as AIM-style structures could be transformed from static classification schemes into live sensing frameworks that support predictive risk detection and continuous mitigation. Key open research directions include empirical evaluation across high-risk domains, reliable mapping from detected signals to mitigation actions, and governance designs that ensure transparency, accountability, and appropriate human control. Ultimately, embedding supervisory agentic monitoring into post-deployment governance could extend the proposed taxonomy from a descriptive framework into a dynamic mechanism that supports continuous, scalable oversight, once the effectiveness of the underlying responses has been evaluated.

\section*{Author Contributions}
Conceptualization, M.R. and D.T.; methodology, E.P., A.G., M.R., L.T.C., I.V. and D.T.; software, E.P., M.R., I.V., D.T.; validation, E.P., A.G.; investigation, E.P., A.G., M.R., L.T.C., I.V. and D.T.; data curation, E.P.; writing—original draft preparation, E.P., A.G. and M.R.; writing—review and editing, E.P., A.G., M.R., I.V. and D.T.; visualization, E.P., M.R., I.V. and D.T.; supervision, M.R., D.T., I.V. and L.T.C; All authors have read and agreed to the published version of the manuscript.

\section*{Funding}
This work is partially financed by the Ministry of Education and Science of the Republic of North Macedonia through the project ``Utilising AI and National Large Language Models to Advance Macedonian Language Capabilities''. This work is also co-funded by the European Union under the Grant Agreement 101263128 (VEZILKA).

\section*{Institutional Review Board Statement}
Not applicable.

\section*{Informed Consent Statement}
Not applicable.

\section*{Data Availability Statement}
The code and dataset for this paper are available on GitHub at: \href{https://github.com/popchanovska/ai-systemic-failures}{github.com/popchanovska/ai-systemic-failures} (accessed on September 21, 2026).

\section*{Acknowledgments}
OpenAI's ChatGPT (model GPT-5.6) was used to a limited extent only for eliminating typos and grammatical improvements during the preparation of the manuscript. The model was not used to generate the scientific content, experimental results, analysis or conclusions.

\section*{Conflicts of Interest}
The authors declare no conflicts of interest.

\newpage

\appendix

\section{Prompts for extracting mitigation actions}
\label{secB1}

\subsection{Version 1}

\textbf{System prompt:}

\noindent You are an AI incident expert specializing in extracting corrective and mitigative actions taken in response to AI incidents. You understand AI incidents, their impacts, and the types of actions organizations and stakeholders take to reduce or contain harm.

\vspace{0.5em}
\noindent \textbf{User prompt:}

\noindent TASK: You will be given a single AI incident description. The incident text is a media or news report describing an AI incident and the actions taken by organizations, individuals, or stakeholders in response.

\noindent DEFINITION OF MITIGATION / CORRECTIVE ACTION: A mitigation or corrective action is any action or process taken in response to an incident to reduce the incident itself or to reduce, limit, contain, or manage the harm caused by the incident. This includes technical, organizational, operational, procedural, legal, policy-based, or public-response actions.

\noindent IMPORTANT CLARIFICATION: Statements of intent, plans, reviews, investigations, warnings, apologies, or condemnations also count as mitigations.

\noindent SCOPE RULES:

\noindent (1) Extract all corrective and mitigative actions explicitly described in the text.  

\noindent (2) Rewrite each action as a full, self-contained sentence so it can be understood without context.  

\noindent (3) Do not infer actions that are not stated. 

\noindent (4) Treat each incident independently.

\noindent OUTPUT FORMAT (STRICT): Return only a list of strings, where each string is a clear explanatory sentence describing a mitigation. If no corrective or mitigative actions are described, return [``No mitigation taken''].

\noindent MITIGATION WRITING RULES:

\noindent (1) Use neutral, factual language.  

\noindent (2) Ensure each sentence is self-contained and readable.  

\noindent (3) Preserve the meaning of the original action, adding minimal clarifying context if needed.  

\noindent (4) Separate multiple actions into distinct sentences.

\noindent AI INCIDENT TEXT:  

\noindent - Text: \{incident\_text\}

\subsection{Version 2}
\textbf{System prompt:}

\noindent You are an AI incident expert specializing in extracting corrective and mitigative actions taken in response to AI incidents. You understand AI incidents, their impacts, and the types of actions organizations and stakeholders take to reduce or contain harm.

\vspace{0.5em}

\noindent \textbf{User prompt:}

\noindent TASK: You will be given a media or news report describing an AI-related incident.

\noindent WHAT TO EXTRACT: Extract all mitigation or corrective actions taken in response to harm, risk, or failure caused by an AI model or AI system.

\noindent WHAT COUNTS AS A MITIGATION: Any explicit response to the AI incident, including technical, organizational, legal, policy, operational, or public-response actions. Apologies, investigations, reviews, enforcement actions, suspensions, removals, and stated plans also count as mitigation.

\noindent RULES:

\noindent (1) Extract only actions that directly respond to the AI-related incident (ignore non-AI causes or background facts).  

\noindent (2) Do not infer actions or motivations not stated in the text.  

\noindent (3) Rewrite each action as one clear, self-contained sentence stating what was done.

\noindent OUTPUT FORMAT: Return only an array of strings. If no AI-related mitigations are described, return [``No mitigation taken''].

\noindent AI INCIDENT TEXT:  

\noindent {incident\_text}

\subsection{Version 3}

\textbf{System prompt:}

\noindent You are an AI incident analysis expert.

\noindent You will be given text from media and news reports describing AI-related incidents. Your task is to extract all explicit post-incident mitigation and corrective actions taken in response to the incident.

\noindent DEFINITION — MITIGATION / CORRECTIVE ACTION: A mitigation or corrective action is any explicit step taken after the incident to respond to, contain, investigate, remediate, reduce harm, acknowledge responsibility, or address consequences caused by an AI system. This includes but is not limited to technical, organizational, operational, legal, policy-based, or public-response actions.

\noindent EXTRACTION RULES:

\noindent (1) Extract only actions explicitly stated in the text.  

\noindent (2) Only extract mitigations from AI-related incidents. If the text does not describe an actual incident, return [``No mitigation taken''].  

\noindent (3) Actions must be taken after the incident; ignore causes or preventive measures taken beforehand.  

\noindent (4) Do not infer or assume actions that are not clearly described.  

\noindent (5) If multiple actions are mentioned together, split them into separate items.  

\noindent (6) Rewrite each action as one clear, self-contained sentence.

\noindent OUTPUT RULES (STRICT):

\noindent Return only a valid JSON array of strings. Do not include explanations, commentary, or extra text. If no mitigation actions are described, return [``No mitigation taken'']. The output must be in English.

\vspace{0.5em}

\noindent \textbf{User prompt:}

\noindent AI INCIDENT TEXT:  

\noindent {incident\_text}

\newpage
\section{Prompt used for classification}
In this prompt, each ``mitigation'' is one record: the complete list of action statements extracted from its article.

\label{secC1}
\subsection{Final version}

\textbf{System prompt:}

\noindent You are an AI incident classification expert.

\noindent Your task is to precisely classify AI incident mitigation actions into predefined mitigation subcategories.

\noindent You must follow the user instructions exactly.

\noindent You must not add explanations, reasoning, commentary, or extra text.

\noindent You must output ONLY valid, machine-parseable JSON.

\noindent You must not invent, infer, generalize, or rename categories.

\noindent All strings must use standard double quotes (\texttt{"}).

\vspace{0.5em}
\noindent \textbf{User prompt:}

\noindent TASK: You will be given a list of AI incident mitigation descriptions.

\noindent Each entry includes an ID and a mitigation description.

\noindent For EACH mitigation, assign one or more mitigation SUBCATEGORY labels.

\vspace{0.5em}
\noindent DEFINITION:

\noindent A ``subcategory'' refers to a specific, fine-grained mitigation type.

\vspace{0.5em}
\noindent ALLOWED SUBCATEGORIES (use ONLY these exact strings):

\begin{itemize}
    \item ``Board Structure \& Oversight''
    \item ``Risk Management''
    \item ``Conflict of Interest Protections''
    \item ``Whistleblower Reporting \& Protection''
    \item ``Safety Decision Frameworks''
    \item ``Environmental Impact Management''
    \item ``Societal Impact Assessment''
    \item ``Model \& Infrastructure Security''
    \item ``Model Alignment''
    \item ``Model Safety Engineering''
    \item ``Content Safety Controls''
    \item ``Testing \& Auditing''
    \item ``Data Governance''
    \item ``Access Management''
    \item ``Staged Deployment''
    \item ``Post-Deployment Monitoring''
    \item ``Incident Response \& Recovery''
    \item ``Incident Investigation''
    \item ``System Documentation''
    \item ``Risk Disclosure''
    \item ``Incident Reporting''
    \item ``Governance Disclosure''
    \item ``Third-Party System Access''
    \item ``User Rights \& Recourse''
    \item ``Training \& Supportive Measures''
    \item ``System \& Feature Restrictions''
    \item ``Usage \& Access Limitations''
    \item ``Denial \& Defensive Based Actions''
    \item ``Court \& Law Enforcement Interventions''
    \item ``Regulatory, Policy \& Legal Mandates''
    \item ``Financial, Economic \& Compensation Remedies''
    \item ``Market Access \& Commercial Restrictions''
\end{itemize}

\vspace{0.5em}
\noindent CLASSIFICATION RULES:

\noindent (1) Assign all subcategories that clearly apply.

\noindent (2) If the mitigation text explicitly states that no action was taken, return [``None''].

\noindent (3) If an action is mentioned but does not match any allowed subcategory, return [``Other''].

\noindent (4) Do not infer actions that are not explicitly stated.

\noindent (5) Do not merge, generalize, or invent subcategories.

\noindent (6) Do not explain your reasoning.

\noindent (7) Do not return any text outside the required JSON structure.

\noindent (8) Subcategory labels must match the allowed list exactly (case-sensitive).

\noindent (9) Each mitigation must be classified independently.

\noindent (10) Assign a maximum of 5 subcategories per mitigation. If more than 5 subcategories could apply, select the 5 most directly and explicitly supported by the mitigation text.

\vspace{0.5em}
\noindent OUTPUT FORMAT (STRICT):

\noindent Return only a single JSON object with this exact structure:

\begin{verbatim}
{
  "results": {
    "<MITIGATION_ID>": {
      "mitigation_text": "<original mitigation text exactly as given>",
      "assigned_subcategories": ["<subcategory>", "..."]
    }
  }
}
\end{verbatim}

\vspace{0.5em}
\noindent MITIGATIONS TO CLASSIFY:

\begin{verbatim}
<mitigations>
{mitigations_block}
</mitigations>
\end{verbatim}

\begin{landscape}

\section{\texorpdfstring{AI Risk Response Taxonomy (RRT)}{AI Risk Response Taxonomy (RRT)} (only new categories and subcategories shown)}
\label{secA1}

\newcolumntype{Z}{>{\raggedright\arraybackslash}p{0.18\linewidth}}
\newcolumntype{Y}{>{\raggedright\arraybackslash}X}

\begin{center}
\begin{tabularx}{\linewidth}{Z Y Y}
\arrayrulecolor{black}\hline
\multicolumn{3}{l}{\textbf{\textcolor{AccentBlue}{Category 3 -- Operational Process Controls}}} \\ \hline
\textbf{Subcategory} & \textbf{Definition} & \textbf{Examples} \\ \hline
\parbox[t]{\hsize}{3.7 \textit{Incident Investigation}} & Post-incident analyses and investigations that assess system behavior, organizational processes, and third-party contributions to identify causes and prevent recurrence. & Root Cause Analysis, Vulnerability Assessment, System Investigation, Vendor Investigation, Internal Reviews, Compliance Audits, Management Investigations \\ \hline
\arrayrulecolor{black}\hline
\multicolumn{3}{l}{\textbf{\textcolor{AccentBlue}{Category 4 -- Transparency \& Accountability Controls}}} \\ \hline
\textbf{Subcategory} & \textbf{Definition} & \textbf{Examples} \\ \hline
\parbox[t]{\hsize}{4.7 \textit{Training \& Supportive Measures}} & Human-centered measures for training and supporting, including help for workers, users of systems, the public, etc. & Workforce retraining programs, Expansion of moderation teams, Redeployment actions, Hiring safety, Education and training, User support and crisis guidance, Professional review, Victim support and counseling, Media literacy and education \\ \hline
\end{tabularx}
\end{center}

\newpage

\begin{center}
\begin{tabularx}{\linewidth}{Z Y Y}
\arrayrulecolor{black}\hline
\multicolumn{3}{l}{\textbf{\textcolor{AccentBlue}{Category 5 -- Corrective \& Restrictive Actions}}} \\ \hline
\textbf{Subcategory} & \textbf{Definition} & \textbf{Examples} \\ \hline
\parbox[t]{\hsize}{5.1 \textit{System \& Feature Restrictions}} & Reactive, incident-triggered or risk-triggered technical or product-level actions that limit, disable, or remove AI system capabilities or features to reduce harm, prevent misuse, or manage risk. & Feature or capability removal (ban/disable), Service or model withdrawal (take offline/sunset), Feature exclusion (preventing specific classes of queries/predictions), Product or model bans, Service suspension, Moratoria on specific features \\ \hline
\parbox[t]{\hsize}{5.2 \textit{Usage \& Access Limitations}} & Reactive, incident-triggered or risk-triggered organisational or operational actions that restrict, pause, or limit the \textit{deployment or use} of an AI system in specific contexts, regions, sectors, or high-risk scenarios. & Halting deployment in a country or region in response to a safety or regulatory concern, Pausing a product rollout following an identified risk, Suspending a high-risk project, Restricting deployment to lower-risk operational sectors, Limiting functionality in specific operational contexts, Project or model suspension for risk mitigation \\ \hline
\end{tabularx}
\end{center}

\newpage

\begin{center}
\begin{tabularx}{\linewidth}{Z Y Y}
\arrayrulecolor{black}\hline
\multicolumn{3}{l}{\textbf{\textcolor{AccentBlue}{Category 6 -- Legal, Regulatory \& Enforcement Actions}}} \\ \hline
\textbf{Subcategory} & \textbf{Definition} & \textbf{Examples} \\ \hline
\parbox[t]{\hsize}{6.1 \textit{Court \& Law Enforcement Interventions}} & AI-incident mitigations imposed by courts and law-enforcement authorities through judicially binding or coercive legal powers. Court-ordered monetary damages receive dual labels with~7.1. & Lawsuits and formal litigation, Cease-and-desist letters (court-issued), Court hearings and orders, Injunctions and consent decrees, \textit{Court-ordered} civil settlement agreements and monetary damages, Criminal investigations, charges, prosecutions and trials, Subpoenas, search warrants and seizures, Law-enforcement actions and dismantling of criminal networks \\ \hline
\parbox[t]{\hsize}{6.2 \textit{Regulatory, Policy \& Legal Mandates}} & AI-incident mitigations imposed by \textit{government agencies or regulatory bodies} to enforce compliance, restrict AI use, or reshape the legal and market environment. & Government and regulatory investigations, Compliance mandates, Enforcement measures, Administrative fines (regulatory), Regulatory directives and statutory amendments, Mandatory reporting obligations (legally imposed), Regulator-driven suspensions and mandatory audits, Government procurement prohibitions, Data-protection authority enforcement, Export controls and blacklists, Legislation and statutory bans \\ \hline
\end{tabularx}
\end{center}

\newpage

\begin{center}
\begin{tabularx}{\linewidth}{Z Y Y}
\arrayrulecolor{black}\hline
\multicolumn{3}{l}{\textbf{\textcolor{AccentBlue}{Category 7 -- Financial, Economic \& Market Controls}}} \\ \hline
\textbf{Subcategory} & \textbf{Definition} & \textbf{Examples} \\ \hline
\parbox[t]{\hsize}{7.1 \textit{Financial, Economic \& Compensation Remedies}} & \textit{Private-sector or voluntarily agreed} financial responses to AI incidents that provide monetary redress or deter future harm. & Voluntary consumer compensation, Restitution frameworks (non-compelled), Pre-litigation or out-of-court settlements, Court-ordered monetary damages (dual-labeled with~6.1), Economic disincentives applied by market actors, Privately negotiated licensing or royalty arrangements as settlement conditions \\ \hline
\parbox[t]{\hsize}{7.2 \textit{Market Access \& Commercial Restrictions}} & Actions that restrict an AI system's \textit{commercial market presence} --- its listing, sale, or procurement in markets. & Removal from app stores or digital marketplaces (delistings), Exclusion from private procurement and contracting frameworks (voluntary), Corporate moratoria on commercial sales of specific AI products, Voluntary restrictions on commercial distribution pending safety review \\ \hline
\end{tabularx}
\end{center}

\newpage

\begin{center}
\begin{tabularx}{\linewidth}{Z Y Y}
\arrayrulecolor{black}\hline
\multicolumn{3}{l}{\textbf{\textcolor{AccentBlue}{Category 8 -- Avoidance \& Denial}}} \\ \hline
\textbf{Subcategory} & \textbf{Definition} & \textbf{Examples} \\ \hline
\parbox[t]{\hsize}{8.1 \textit{Denial \& Defensive-Based Actions}} & Explicit and public responses by the party primarily responsible for an AI incident in which the actor rejects responsibility, denies the existence of harm, or declines to take corrective action. \textit{Avoidance}: non-response, material delay, or deliberate non-disclosure of a known harm. \textit{Denial}: active public rejection of responsibility or the existence of harm. May be assigned together with remediation labels (rule d, Section~\ref{data_and_methodology}). Note: Track B response (institutional posturing), not harm-reducing mitigation; see Section~\ref{dualtrack}. & Formal denial of responsibility or intent, Public statements asserting full legal compliance in lieu of remediation, Refusal to comply with removal or disassociation requests, Reiteration of legal or policy defences in response to complaints, Non-response or deliberate delay in acknowledging a known AI-related harm \\ \hline
\end{tabularx}
\end{center}
\end{landscape}

\section{\texorpdfstring{Per-subcategory chance-corrected agreement}{Per-subcategory chance-corrected agreement}}
\label{secD1}
Tables~\ref{tab:kappa-a1a2}--\ref{tab:kappa-llma2} report the agreement for each of the 32 subcategories in the validation sample of 300 records (Section~\ref{sec:catvalidation}). For each subcategory, $a$ is the number of records to which both label sets assign it, $b$ the number assigned it only by the first label set, $c$ only by the second, and $d$ by neither. $P_o = (a+d)/N$ is the observed agreement. $P_e$ is the chance agreement, computed from each rater's own label rate for Cohen's $\kappa$ and from the pooled rate of both raters for Fleiss' $\kappa$, and $\kappa = (P_o - P_e)/(1 - P_e)$.

\begin{table}[H]
\centering
\caption{Per-subcategory agreement, annotator 1 (first) vs annotator 2 (second) ($N = 300$ records).}
\label{tab:kappa-a1a2}
\scriptsize
\setlength{\tabcolsep}{3pt}
\resizebox{\linewidth}{!}{%
\begin{tabular}{@{} l r r r r c c c c c @{}}
\toprule
\textbf{Subcategory} & $a$ & $b$ & $c$ & $d$ & $P_o$ & $P_e$ (Cohen) & $\kappa$ (Cohen) & $P_e$ (Fleiss) & $\kappa$ (Fleiss) \\
\midrule
Access Management & 34 & 5 & 18 & 243 & 0.923 & 0.742 & 0.703 & 0.743 & 0.702 \\
Board Structure \& Oversight & 8 & 4 & 6 & 282 & 0.967 & 0.917 & 0.598 & 0.917 & 0.598 \\
Conflict of Interest Protections & 1 & 2 & 7 & 290 & 0.970 & 0.964 & 0.170 & 0.964 & 0.167 \\
Content Safety Controls & 49 & 12 & 18 & 221 & 0.900 & 0.664 & 0.702 & 0.664 & 0.702 \\
Court \& Law Enforcement Interventions & 50 & 5 & 3 & 242 & 0.973 & 0.705 & 0.910 & 0.705 & 0.910 \\
Data Governance & 44 & 8 & 13 & 235 & 0.930 & 0.703 & 0.765 & 0.703 & 0.765 \\
Denial \& Defensive Based Actions & 20 & 12 & 10 & 258 & 0.927 & 0.815 & 0.604 & 0.815 & 0.604 \\
Environmental Impact Management & 10 & 0 & 0 & 290 & 1.000 & 0.936 & 1.000 & 0.936 & 1.000 \\
Financial, Economic \& Compensation Remedies & 27 & 4 & 2 & 267 & 0.980 & 0.820 & 0.889 & 0.820 & 0.889 \\
Governance Disclosure & 36 & 10 & 11 & 243 & 0.930 & 0.738 & 0.733 & 0.738 & 0.733 \\
Incident Investigation & 46 & 3 & 7 & 244 & 0.967 & 0.718 & 0.882 & 0.718 & 0.882 \\
Incident Reporting & 45 & 11 & 12 & 232 & 0.923 & 0.694 & 0.749 & 0.694 & 0.749 \\
Incident Response \& Recovery & 47 & 4 & 8 & 241 & 0.960 & 0.709 & 0.863 & 0.709 & 0.863 \\
Market Access \& Commercial Restrictions & 27 & 3 & 1 & 269 & 0.987 & 0.825 & 0.924 & 0.825 & 0.924 \\
Model \& Infrastructure Security & 32 & 2 & 5 & 261 & 0.977 & 0.791 & 0.888 & 0.791 & 0.888 \\
Model Alignment & 12 & 0 & 7 & 281 & 0.977 & 0.902 & 0.763 & 0.902 & 0.762 \\
Model Safety Engineering & 22 & 5 & 8 & 265 & 0.957 & 0.828 & 0.748 & 0.828 & 0.748 \\
Post-Deployment Monitoring & 27 & 2 & 13 & 258 & 0.950 & 0.796 & 0.755 & 0.796 & 0.754 \\
Regulatory, Policy \& Legal Mandates & 43 & 5 & 9 & 243 & 0.953 & 0.722 & 0.832 & 0.722 & 0.832 \\
Risk Disclosure & 42 & 6 & 12 & 240 & 0.940 & 0.718 & 0.788 & 0.718 & 0.787 \\
Risk Management & 35 & 4 & 18 & 243 & 0.927 & 0.739 & 0.719 & 0.740 & 0.718 \\
Safety Decision Frameworks & 15 & 6 & 13 & 266 & 0.937 & 0.850 & 0.579 & 0.850 & 0.578 \\
Societal Impact Assessment & 20 & 9 & 18 & 253 & 0.910 & 0.801 & 0.547 & 0.802 & 0.546 \\
Staged Deployment & 17 & 4 & 6 & 273 & 0.967 & 0.864 & 0.755 & 0.864 & 0.755 \\
System \& Feature Restrictions & 36 & 3 & 20 & 241 & 0.923 & 0.732 & 0.714 & 0.733 & 0.712 \\
System Documentation & 9 & 2 & 8 & 281 & 0.967 & 0.911 & 0.626 & 0.911 & 0.625 \\
Testing \& Auditing & 43 & 0 & 14 & 243 & 0.953 & 0.721 & 0.833 & 0.722 & 0.832 \\
Third-Party System Access & 19 & 4 & 20 & 257 & 0.920 & 0.813 & 0.572 & 0.815 & 0.568 \\
Training \& Supportive Measures & 44 & 3 & 14 & 239 & 0.943 & 0.711 & 0.804 & 0.711 & 0.804 \\
Usage \& Access Limitations & 31 & 5 & 20 & 244 & 0.917 & 0.751 & 0.666 & 0.752 & 0.664 \\
User Rights \& Recourse & 30 & 6 & 12 & 252 & 0.940 & 0.774 & 0.735 & 0.774 & 0.735 \\
Whistleblower Reporting \& Protection & 10 & 3 & 1 & 286 & 0.987 & 0.923 & 0.826 & 0.923 & 0.826 \\
\bottomrule
\end{tabular}}
\end{table}

\begin{table}[H]
\centering
\caption{Per-subcategory agreement, LLM (first) vs annotator 1 (second) ($N = 300$ records).}
\label{tab:kappa-llma1}
\scriptsize
\setlength{\tabcolsep}{3pt}
\resizebox{\linewidth}{!}{%
\begin{tabular}{@{} l r r r r c c c c c @{}}
\toprule
\textbf{Subcategory} & $a$ & $b$ & $c$ & $d$ & $P_o$ & $P_e$ (Cohen) & $\kappa$ (Cohen) & $P_e$ (Fleiss) & $\kappa$ (Fleiss) \\
\midrule
Access Management & 34 & 11 & 5 & 250 & 0.947 & 0.759 & 0.779 & 0.759 & 0.779 \\
Board Structure \& Oversight & 11 & 2 & 1 & 286 & 0.990 & 0.920 & 0.875 & 0.920 & 0.875 \\
Conflict of Interest Protections & 1 & 5 & 2 & 292 & 0.977 & 0.970 & 0.212 & 0.970 & 0.210 \\
Content Safety Controls & 48 & 2 & 13 & 237 & 0.950 & 0.698 & 0.835 & 0.698 & 0.834 \\
Court \& Law Enforcement Interventions & 49 & 1 & 6 & 244 & 0.977 & 0.711 & 0.919 & 0.711 & 0.919 \\
Data Governance & 46 & 4 & 6 & 244 & 0.967 & 0.718 & 0.882 & 0.718 & 0.882 \\
Denial \& Defensive Based Actions & 15 & 10 & 17 & 258 & 0.910 & 0.828 & 0.477 & 0.828 & 0.477 \\
Environmental Impact Management & 10 & 0 & 0 & 290 & 1.000 & 0.936 & 1.000 & 0.936 & 1.000 \\
Financial, Economic \& Compensation Remedies & 29 & 3 & 2 & 266 & 0.983 & 0.812 & 0.911 & 0.812 & 0.911 \\
Governance Disclosure & 40 & 10 & 6 & 244 & 0.947 & 0.731 & 0.802 & 0.731 & 0.802 \\
Incident Investigation & 46 & 4 & 3 & 247 & 0.977 & 0.724 & 0.915 & 0.724 & 0.915 \\
Incident Reporting & 47 & 3 & 9 & 241 & 0.960 & 0.709 & 0.863 & 0.709 & 0.863 \\
Incident Response \& Recovery & 45 & 5 & 6 & 244 & 0.963 & 0.720 & 0.869 & 0.720 & 0.869 \\
Market Access \& Commercial Restrictions & 27 & 1 & 3 & 269 & 0.987 & 0.825 & 0.924 & 0.825 & 0.924 \\
Model \& Infrastructure Security & 31 & 2 & 3 & 264 & 0.983 & 0.802 & 0.916 & 0.802 & 0.916 \\
Model Alignment & 12 & 3 & 0 & 285 & 0.990 & 0.914 & 0.884 & 0.914 & 0.884 \\
Model Safety Engineering & 22 & 2 & 5 & 271 & 0.977 & 0.844 & 0.850 & 0.844 & 0.850 \\
Post-Deployment Monitoring & 28 & 10 & 1 & 261 & 0.963 & 0.801 & 0.816 & 0.802 & 0.815 \\
Regulatory, Policy \& Legal Mandates & 44 & 6 & 4 & 246 & 0.967 & 0.727 & 0.878 & 0.727 & 0.878 \\
Risk Disclosure & 43 & 3 & 5 & 249 & 0.973 & 0.736 & 0.899 & 0.736 & 0.899 \\
Risk Management & 35 & 10 & 4 & 251 & 0.953 & 0.759 & 0.806 & 0.759 & 0.806 \\
Safety Decision Frameworks & 15 & 0 & 6 & 279 & 0.980 & 0.887 & 0.823 & 0.887 & 0.823 \\
Societal Impact Assessment & 17 & 1 & 12 & 270 & 0.957 & 0.855 & 0.701 & 0.856 & 0.700 \\
Staged Deployment & 19 & 7 & 2 & 272 & 0.970 & 0.855 & 0.792 & 0.856 & 0.792 \\
System \& Feature Restrictions & 37 & 13 & 2 & 248 & 0.950 & 0.747 & 0.803 & 0.747 & 0.802 \\
System Documentation & 9 & 6 & 2 & 283 & 0.973 & 0.917 & 0.679 & 0.917 & 0.678 \\
Testing \& Auditing & 43 & 7 & 0 & 250 & 0.977 & 0.738 & 0.911 & 0.738 & 0.911 \\
Third-Party System Access & 21 & 11 & 2 & 266 & 0.957 & 0.833 & 0.740 & 0.833 & 0.740 \\
Training \& Supportive Measures & 45 & 5 & 2 & 248 & 0.977 & 0.729 & 0.914 & 0.729 & 0.914 \\
Usage \& Access Limitations & 31 & 7 & 5 & 257 & 0.960 & 0.784 & 0.815 & 0.784 & 0.815 \\
User Rights \& Recourse & 30 & 5 & 6 & 259 & 0.963 & 0.791 & 0.824 & 0.791 & 0.824 \\
Whistleblower Reporting \& Protection & 11 & 0 & 2 & 287 & 0.993 & 0.923 & 0.913 & 0.923 & 0.913 \\
\bottomrule
\end{tabular}}
\end{table}

\begin{table}[H]
\centering
\caption{Per-subcategory agreement, LLM (first) vs annotator 2 (second) ($N = 300$ records).}
\label{tab:kappa-llma2}
\scriptsize
\setlength{\tabcolsep}{3pt}
\resizebox{\linewidth}{!}{%
\begin{tabular}{@{} l r r r r c c c c c @{}}
\toprule
\textbf{Subcategory} & $a$ & $b$ & $c$ & $d$ & $P_o$ & $P_e$ (Cohen) & $\kappa$ (Cohen) & $P_e$ (Fleiss) & $\kappa$ (Fleiss) \\
\midrule
Access Management & 43 & 2 & 9 & 246 & 0.963 & 0.729 & 0.865 & 0.729 & 0.865 \\
Board Structure \& Oversight & 10 & 3 & 4 & 283 & 0.977 & 0.914 & 0.729 & 0.914 & 0.729 \\
Conflict of Interest Protections & 6 & 0 & 2 & 292 & 0.993 & 0.954 & 0.854 & 0.954 & 0.854 \\
Content Safety Controls & 49 & 1 & 18 & 232 & 0.937 & 0.684 & 0.799 & 0.686 & 0.798 \\
Court \& Law Enforcement Interventions & 50 & 0 & 3 & 247 & 0.990 & 0.716 & 0.965 & 0.716 & 0.965 \\
Data Governance & 47 & 3 & 10 & 240 & 0.957 & 0.707 & 0.852 & 0.707 & 0.852 \\
Denial \& Defensive Based Actions & 23 & 2 & 7 & 268 & 0.970 & 0.833 & 0.820 & 0.833 & 0.820 \\
Environmental Impact Management & 10 & 0 & 0 & 290 & 1.000 & 0.936 & 1.000 & 0.936 & 1.000 \\
Financial, Economic \& Compensation Remedies & 29 & 3 & 0 & 268 & 0.990 & 0.817 & 0.945 & 0.817 & 0.945 \\
Governance Disclosure & 45 & 5 & 2 & 248 & 0.977 & 0.729 & 0.914 & 0.729 & 0.914 \\
Incident Investigation & 50 & 0 & 3 & 247 & 0.990 & 0.716 & 0.965 & 0.716 & 0.965 \\
Incident Reporting & 47 & 3 & 10 & 240 & 0.957 & 0.707 & 0.852 & 0.707 & 0.852 \\
Incident Response \& Recovery & 50 & 0 & 5 & 245 & 0.983 & 0.711 & 0.942 & 0.711 & 0.942 \\
Market Access \& Commercial Restrictions & 28 & 0 & 0 & 272 & 1.000 & 0.831 & 1.000 & 0.831 & 1.000 \\
Model \& Infrastructure Security & 33 & 0 & 4 & 263 & 0.987 & 0.794 & 0.935 & 0.794 & 0.935 \\
Model Alignment & 15 & 0 & 4 & 281 & 0.987 & 0.893 & 0.875 & 0.893 & 0.875 \\
Model Safety Engineering & 24 & 0 & 6 & 270 & 0.980 & 0.836 & 0.878 & 0.836 & 0.878 \\
Post-Deployment Monitoring & 37 & 1 & 3 & 259 & 0.987 & 0.774 & 0.941 & 0.774 & 0.941 \\
Regulatory, Policy \& Legal Mandates & 49 & 1 & 3 & 247 & 0.987 & 0.718 & 0.953 & 0.718 & 0.953 \\
Risk Disclosure & 44 & 2 & 10 & 244 & 0.960 & 0.722 & 0.856 & 0.722 & 0.856 \\
Risk Management & 44 & 1 & 9 & 246 & 0.967 & 0.726 & 0.878 & 0.727 & 0.878 \\
Safety Decision Frameworks & 15 & 0 & 13 & 272 & 0.957 & 0.866 & 0.677 & 0.867 & 0.674 \\
Societal Impact Assessment & 18 & 0 & 20 & 262 & 0.933 & 0.829 & 0.611 & 0.831 & 0.606 \\
Staged Deployment & 22 & 4 & 1 & 273 & 0.983 & 0.850 & 0.889 & 0.850 & 0.889 \\
System \& Feature Restrictions & 46 & 4 & 10 & 240 & 0.953 & 0.709 & 0.840 & 0.709 & 0.840 \\
System Documentation & 14 & 1 & 3 & 282 & 0.987 & 0.899 & 0.868 & 0.899 & 0.868 \\
Testing \& Auditing & 49 & 1 & 8 & 242 & 0.970 & 0.707 & 0.898 & 0.707 & 0.898 \\
Third-Party System Access & 30 & 2 & 9 & 259 & 0.963 & 0.791 & 0.825 & 0.791 & 0.824 \\
Training \& Supportive Measures & 49 & 1 & 9 & 241 & 0.967 & 0.704 & 0.887 & 0.705 & 0.887 \\
Usage \& Access Limitations & 38 & 0 & 13 & 249 & 0.957 & 0.746 & 0.829 & 0.747 & 0.828 \\
User Rights \& Recourse & 35 & 0 & 7 & 258 & 0.977 & 0.776 & 0.896 & 0.776 & 0.896 \\
Whistleblower Reporting \& Protection & 10 & 1 & 1 & 288 & 0.993 & 0.929 & 0.906 & 0.929 & 0.906 \\
\bottomrule
\end{tabular}}
\end{table}

\newpage

\bibliographystyle{unsrt}
\bibliography{sn-bibliography}

\end{document}